\documentclass[journal]{IEEEtran}

\usepackage[ left=0.625in, top=0.75in, right=0.75in, bottom=1.22in, headsep=0.25in, headheight=20pt, heightrounded, letterpaper] {geometry}
\usepackage[utf8]{inputenc}
\usepackage{lipsum}
\usepackage{comment}
\usepackage{xfrac}  

\usepackage{lipsum,microtype}
\usepackage{authblk}
\usepackage[ruled,vlined]{algorithm2e}
\usepackage{amsmath, nccmath}
\usepackage{geometry}
\usepackage{tabularx}
\usepackage{booktabs}
\usepackage{setspace}
\usepackage[labelfont=bf,format=plain,justification=raggedright,singlelinecheck=true]{caption}
\usepackage{mathtools}
\usepackage{amssymb}
\usepackage{caption}
\usepackage{svg}
\usepackage{xcolor}
\usepackage{subcaption}
\usepackage{graphicx}
\graphicspath{{./Fig/}} 
\newcommand*{\Scale}[2][4]{\scalebox{#1}{$#2$}}%
\usepackage[ruled,vlined]{algorithm2e}
\usepackage{cite}
\newcounter{mycounter}
\usepackage{subfig} 
\usepackage{fancyhdr}
\usepackage{textcomp}
\usepackage{blindtext}
\usepackage{optidef}
\usepackage{subcaption} 
\usepackage{ragged2e} 
\usepackage{multirow}
\usepackage{longtable}
\usepackage{pifont}
\usepackage{cuted} 
\usepackage{stfloats}

\usepackage{fancyhdr}
\usepackage{textcomp}
\usepackage{blindtext}

\title{GenAI-enabled Residual Motion Estimation for Energy-Efficient Semantic Video Communication}
\author{Shavbo Salehi\IEEEauthorrefmark{1}, \textit{Student Member, IEEE}, Pedro Enrique Iturria-Rivera\IEEEauthorrefmark{2}, Medhat Elsayed\IEEEauthorrefmark{2}, Majid Bavand\IEEEauthorrefmark{2}, Yigit Ozcan\IEEEauthorrefmark{2}, Melike Erol-Kantarci\IEEEauthorrefmark{1}, \textit{Fellow, IEEE}\\

\IEEEauthorblockA{\IEEEauthorrefmark{1}\textit{School of Electrical Engineering and Computer Science, University of Ottawa, Ottawa, Canada}}\\
\IEEEauthorblockA{\IEEEauthorrefmark{2}\textit{Ericsson Canada Inc., Ottawa, Canada}}\\
\IEEEauthorblockA{Emails: \{ssale038, melike.erolkantarci\}@uottawa.ca, \\\{pedro.iturria.rivera,medhat.elsayed,majid.bavand, yigit.ozcan\}@ericsson.com}}

\date{}
\begin{document}

\twocolumn

\maketitle
\begin{abstract}

Semantic communication addresses the limitations of the Shannon paradigm by focusing on transmitting meaning rather than exact representations, thereby reducing unnecessary resource consumption. This is particularly beneficial for video, which dominates network traffic and demands high bandwidth and power, making semantic approaches ideal for conserving resources while maintaining quality. In this paper, we propose a Predictability-aware and Entropy-adaptive Neural Motion Estimation (PENME) method to address challenges related to high latency, high bitrate, and power consumption in video transmission. PENME makes per-frame decisions to select a residual motion extraction model, convolutional neural network, vision transformer, or optical flow, using a five-step policy based on motion strength, global motion consistency, peak sharpness, heterogeneity, and residual error. The residual motions are then transmitted to the receiver, where the frames are reconstructed via motion-compensated updates. Next, a selective diffusion-based refinement, the Latent Consistency Model (LCM-4), is applied on frames that trigger refinement due to low predictability or large residuals, while predictable frames skip refinement. PENME also allocates radio resource blocks with awareness of residual motion and channel state, reducing power consumption and bandwidth usage while maintaining high semantic similarity. Our simulation results on the Vimeo90K dataset demonstrate that the proposed PENME method handles various types of video, outperforming traditional communication, hybrid, and adaptive bitrate semantic communication techniques, achieving 40\% lower latency, 90\% less transmitted data, and 35\% higher throughput. For semantic communication metrics, PENME improves peak signal-to-noise ratio (PSNR) by about $40\%$, increases multi-scale structural similarity (MS-SSIM) by roughly $19\%$, and reduces learned perceptual image patch similarity (LPIPS) by nearly $35\%$, compared with the baseline methods.

\end{abstract}

\begin{IEEEkeywords}
Diffusion-Based Frame Refinement, Entropy-Aware Feature Selection, Semantic Communication, Residual Motion Estimation, Video Transmission Efficiency.\vspace{-0.3cm}
\end{IEEEkeywords}

\section{Introduction}

Semantic communication has emerged as a key enabler for beyond-5G and 6G networks, addressing the limitations of conventional communication systems by transmitting the essential meaning of data rather than reconstructing every bit \cite{strinati2024goal,sagduyu2024will,fernandes2024semantic}. Unlike Shannon’s paradigm, which optimizes the accuracy of bit transmission, semantic communication prioritizes the efficiency and reliability of information exchange by extracting and transmitting features relevant to the task \cite{qin2021semantic}. This paradigm shift reduces redundancy, enhances robustness against noise and interference, and enables low-latency communication, making it practical for high bandwidth applications \cite{chaccour2024less}. Although semantic communication has been investigated for text \cite{KG-LLM} and images \cite{zhang2023optimization}, video remains the most challenging modality due to its massive size, strong temporal correlations, and stringent quality of service demands \cite{guo2025videoqa,aloudat2025metaverse}. Furthermore, ensuring efficient and reliable video delivery is critical to realize the potential of semantic communication in next-generation networks, as pushing more bits through the channel has proven to be an unsustainable approach \cite{fernandes2024semantic, chaccour2024less}.

As a solution for transmitting videos in semantic-based communications, encoders have been developed to capture not only objects and scenes, but also the motion dynamics and temporal relationships between frames \cite{li2024video}. Attribute–relation-based representations distill the frames into objects, properties, and spatial relationships, thus reducing data while preserving meaning \cite{du2025object}. However, aggressive semantic compression can lead to information loss, omitting subtle contextual signals whose absence degrades both user experience and task performance \cite{jiang2022wireless}. Then, adaptive approaches \cite{wang2023adaptive} have been considered as a viable solution. In \cite{gong2023adaptive}, an adaptive bitrate video semantic communication (ABRVSC) method was proposed, which adapts the bitrates of video semantic data using a swin transformer–based semantic codec according to network conditions, aiming to optimize both transmission delay and semantic accuracy. Task-oriented frameworks such as task-oriented adaptive semantic transmission (TOAST) further extend this idea by aligning communication strategies with task goals and optimizing semantic accuracy under dynamic conditions by swin transformer-based architecture \cite{yun2025toast}. Despite various utilization of adaptive codecs, they have remained reliant on motion estimation and prediction-based coding, which become computationally expensive and degrade under complex or unpredictable motion patterns \cite{VVCsamarathunga2024semantic}. 

To overcome the limitations of adaptive methods, recent research has shifted toward generative approaches that avoid explicit motion modeling and instead use high-level representations \cite{yin2025generative}. By utilizing generative and multimodal methods, high-level features such as text, audio, or latent representations are transmitted instead of full video streams \cite{yan2025semantic}. Then in the receiver (Rx) side, the videos are reconstructed using generative adversarial networks (GANs)- or diffusion-based decoders \cite{chen2024generative, tong2024multimodal}. Diffusion-based codecs improve realism and denoising \cite{wu2024cddm, pei2025latent}, but their iterative sampling makes them computationally intensive and less suitable for latency-sensitive applications. Errors in segmentation or keypoint extraction propagate to reconstruction, making these methods fragile when the motion is rapid or unpredictable. Furthermore, most existing systems depend on large pre-trained models that cannot adapt online to channel variations, resulting in vulnerability to fading, packet loss, and segmentation errors \cite{jiang2022wireless}. These shortcomings collectively highlight the need for a framework that balances motion fidelity, semantic preservation, adaptability, and computational feasibility.

To address these challenges, we propose the predictability- and entropy-aware neural motion estimation (PENME) framework, a semantic video communication system that unifies adaptive motion modeling with generative refinement. On the transmitter (Tx) side of PENME, normalized motion-related signals are evaluated to dynamically select the most suitable motion estimation strategy. For this aim, PENME derives five bounded motion signals from each pair of frames, including motion strength, global shift consistency, peak confidence of that alignment, local motion heterogeneity from block matching, and the residual mismatch after removing a robust global shift, all normalized to a common scale for fair comparison. These signals are then used in an interpretable score that routes each segment to the most suitable estimator: optical flow for strong, coherent motion; a vision transformer (ViT) for complex or irregular motion; and convolutional neural networks (CNN) for weak, homogeneous motion. This adaptive selection ensures efficient resource allocation while maintaining motion accuracy. Then on the Rx side, the frames are reconstructed, and a conditional latent consistency refinement module (LCM-4) refines the latent transmitted motion projecting noisy or incomplete motion vectors onto a temporally coherent manifold. This approach improves frame reconstruction and mitigates the effects of fading on frames, packet loss, and quantization. Unlike full generative decoders, PENME applies diffusion only as a few-step refinement in the latent space, which reduces sampling cost and decoder load. As a result, PENME improves semantic fidelity, enhances robustness under dynamic and noisy conditions, and reduces redundancy in video transmission, bridging the gap left by prior approaches. The key novelties and contributions of this paper are summarized as follows:

\begin{itemize}
    \item \textbf{Predictability- and Entropy-Aware Motion Estimation:} We introduce a dynamic selection mechanism that predominantly assigns CNNs for simple and homogeneous motion, optical flow for segments with strong but globally consistent motion, and ViTs for frames with highly irregular or complex motions. This adaptive strategy balances efficiency with accuracy in diverse motion conditions.
    
    \item \textbf{Conditional Diffusion-based refinement:} PENME uses a conditional diffusion module to refine residual motion degraded by noise, packet loss, or unpredictable motion. With a few-step latent sampling for real-time operation, it avoids the cost of full diffusion while preserving bandwidth and maintaining high-quality reconstruction under challenging conditions and complex frames.

    \item \textbf{Edge-Assisted and Deployment-Oriented Design:} PENME uses pre-trained feature extractors and shifts heavy computation to the edge, reducing the Tx’s load. By offloading diffusion-based refinement and reconstruction, it also keeps Rx overhead low while remaining practical for wireless video deployment.
    
    \item \textbf{Comprehensive Evaluation:} We benchmark PENME against state-of-the-art semantic and neural video codecs using both network-level metrics (throughput, latency, MSE, PDR, and BER) and semantic quality metrics (PSNR, MS-SSIM, and LPIPS). Across various channel conditions, PENME consistently delivers higher visual quality, improved robustness, and lower bitrate compared to the hybrid and ABRVSC methods.
\end{itemize}

The remainder of this paper is organized as follows. Section \ref{section:LR} presents a review of the literature covering recent advances in semantic communication and highlights key limitations in existing video transmission methods. Section \ref{section:system} introduces the system model, detailing the network architecture, data transmission, and the semantic-aware communication framework. Section \ref{section:performance} describes the proposed PENME method, including its motion extraction strategy and diffusion-based refinement. Section \ref{section:results} provides the experimental setup and evaluation results, demonstrating PENME’s performance compared to ABRVSC, traditional, and hybrid baselines under various network and video conditions. Finally, Section \ref{section:conclusion} concludes the paper by summarizing the contributions and outlining potential directions for future work.

\section{Literature Review}
\label{section:LR}

\begin{table*}[ht]
\centering
\caption{Literature review of semantic video communication methods.}
\label{table:LR}

\begingroup
\fontsize{9pt}{14pt}\selectfont
\renewcommand{\arraystretch}{0.8}
\setlength{\tabcolsep}{3pt}

\begin{tabular}{|
  >{\raggedright\arraybackslash}p{1.1cm} |
  >{\raggedright\arraybackslash}p{3.1cm} |
  >{\raggedright\arraybackslash}p{2cm} | 
  >{\raggedright\arraybackslash}p{1.7cm} | 
  >{\raggedright\arraybackslash}p{2cm} | 
  >{\raggedright\arraybackslash}p{3.2cm} | 
  >{\raggedright\arraybackslash}p{2.7cm} | 
}
\hline
\textbf{Ref} & \textbf{Method Summary} & \textbf{Motion Handling} & \textbf{Robustness to Noise} & \textbf{Complexity} & \textbf{Advantages} & \textbf{Limitations} \\ \hline

\cite{yan2025semantic} & Latent diffusion with I-frame compression and P/B metadata & Frame interpolation & High & Moderate–High & Efficient bandwidth; high QoE; streaming-friendly & Diffusion decoding overhead \\ \hline

\cite{wu2024cddm} & Channel denoising diffusion (physical-layer JSCC) & None & Very High & High & Strong noise suppression & No motion modeling; heavy iterative decoding \\ \hline

\cite{pei2025latent} & Latent diffusion + adversarial encoder & Latent temporal prediction & High & Moderate & Robust to OOD/channel noise; low latency decoding & Limited explicit motion modeling \\ \hline

\cite{yun2025toast} & Task-oriented semantic adaptation (multi-task + RL) & Implicit via task signals & High & High & Adaptive; task-aware fidelity & Complex; heavy decoding \\ \hline

\cite{samarathunga2024semantic} & Hybrid semantic codec + VVC & Standard codec prediction & Low–Medium & Moderate–High & Codec-compatible; efficient compression & Not robust to noisy channels \\ \hline

\cite{chen2024generative} & M3E-VSC: multimodal GAN-based fusion & Multi-frame semantic alignment & High & High & Strong semantic recovery; multimodal context & GAN instability; heavy models \\ \hline

\cite{tong2024multimodal} & Wav2Vid: audio-to-video GAN & Implicit lip-motion synthesis & Medium & High & Bandwidth saving; realistic lip-sync & Limited application scope; GAN overhead \\ \hline

\cite{li2024video} & MOE-CVE: object-based contextual encoding & CNN-based motion estimation & Medium & High & Bitrate reduction via object focus & Depends on segmentation; weak for complex motion \\ \hline

\cite{han2024motion} & DL-based Motion-aware frame interpolation & Explicit motion interpolation & Low & High & High-quality skipped frames prediction & High compute; poor noise robustness \\ \hline

\cite{huang2019comyco} & Comyco: ABR via imitation learning & Codec-level only & Low & Low & QoE optimization; practical baseline & No semantic compression \\ \hline

\textbf{PENME} & \textbf{Adaptive semantic codec selection + diffusion refinement} & \textbf{Adaptive motion extraction} & \textbf{Very High} & \textbf{Moderate (model switching)} & \textbf{QoE optimization; bandwidth saving; semantic preservation} & \textbf{Training cost; edge computation} \\ \hline

\end{tabular}
\endgroup
\end{table*}

In recent years, semantic communication has gained attention as an effective approach for overcoming the limitations of traditional wireless communication systems. Semantic communication is applicable in various scenarios, including smart cities \cite{liang2024generative}, robotics \cite{zeng2024knowledge}, and transportation \cite{wan2024semantic}, each of which requires different data modalities \cite{uysal2022semantic}. Among these, text is typically the simplest to process, as it has a small payload and its meaning can be extracted straightforwardly \cite{KG-LLM}. Compared to text, images generally have larger data sizes, requiring the allocation of more resource blocks (RB) and more complex semantic extraction methods \cite{erdemir2023generative}. Although semantic techniques have seen extensive development for text and images, video poses unique difficulties due to its scale and temporal complexity \cite{jiang2022wireless,guo2025videoqa}. Furthermore, video streaming constitutes a dominant share of the Internet traffic, and requires more efficient transmission methods to maintain quality under bandwidth constraints and stringent latency requirements \cite{guo2025videoqa}. In general, previous work on semantic video communication to address its requirements and challenges has been categorized into three themes: (i) modality- and generation-aware semantic encoding, (ii) motion-aware and object-centric semantic encoding, and (iii) network resource-aware transmission and adaptation.

Multimodal based methods have been used to minimize redundant video frame transmission by extracting high-level semantic data (e.g. text, audio, or low-dimensional features) on the Tx side, and reconstructing the video using generative models at the Rx side. This approach can decrease bandwidth usage while preserving perceptual quality \cite{chen2024generative,tong2024multimodal}. In \cite{chen2024generative}, a multimodal mutual enhancement video semantic communication system (M3E-VSC) was proposed that transmits text as the main carrier of video content from key frames that convey the meaning of scenes, then on the Rx side a generative model reconstructs the video frames, improving bandwidth efficiency and remaining robust at low signal-to-noise ratio (SNR). In \cite{tong2024multimodal}, a wave-to-video method (Wav2Vid) was proposed that focuses on audio-driven video generation and relies on a GAN-based decoder to synthesize talking-head videos from the audio stream. In \cite{tian2025synchronous}, the synchronous multimodal system (SyncSC) was proposed, which transmits 3D morphable model (3DMM) coefficients for facial video and text for speech, and uses generative decoders in the receiving region, including 3DMM image rendering and neural speech synthesis, to reconstruct aligned audio and visual streams. 


Despite their promise, generative semantic video methods face notable challenges. Using text \cite{chen2024generative} as a video representation can lead to a semantic loss of fine details and fail to reproduce subtle features, degrading downstream tasks such as object recognition and reducing perceptual quality \cite{jiang2022wireless}. Audio-driven systems \cite{tong2024multimodal} can retain more realism under the assumption of relatively static scenes, but rapid scene changes or unpredictable motions can break this assumption. It should be noted that the reliance of these methods on large neural models on both the Tx and Rx sides increases complexity, which can hinder real-time deployment on resource-constrained devices or edge platforms. Recent motion-aware frame interpolation codecs (e.g., motion-aware video frame interpolation (MA-VFI) \cite{han2024motion}, motion-aware diffusion for video frame interpolation (MADiff) \cite{ma2024madiff}) show strong temporal coherence under complex motion. However, these methods are highly dependent on accurate neighboring frames and dense motion estimates, and are sensitive to rapid changes that lead to low tolerability to packet loss and latency \cite{ma2024madiff}. Researchers have also explored latent diffusion–based video compression \cite{wu2025improved,wang2025lavie,danier2024ldmvfi}, which learns compact latent representations and preserves perceptual quality. Its limitations include multi-stage decoders with several denoising steps, high memory and computational requirements at the Rx, slow bitrate adaptation, and an assumption of stable transport. As a solution, advanced vector-quantized diffusion extensions (VQ-Diffusion) \cite{nguyen2024improving,yang2023improving} combine vector quantization with diffusion priors to capture temporal dependencies, offering a solution for low-bitrate semantic video communication. VQ-Diffusion systems, while achieving strong perceptual quality for offline generation or interpolation, carry large codebooks, produce code-index streams that are unstable to bit errors, add delay through temporal modules, and lack built-in channel-aware rate control. 

Another research direction focuses on extracting and transmitting only the most relevant parts of each frame or motion, rather than the entire pixel array, to exploit temporal and spatial redundancies in video. In \cite{jiang2022wireless}, a semantic video conferencing (SVC) method was proposed that detects facial landmarks to represent a speaker’s facial motion and transmits only these coordinates. Similarly, \cite{hu2022semantic} employed semantic neural rendering for talking-head video by sending compact facial semantics (e.g., landmarks/coefficients) and reconstructing frames at the Rx via a neural renderer, achieving ultra–low bitrate conferencing. Beyond human faces, \cite{du2025object} proposed an object–attribute–relation (OAR) representation for general scenes, distilling frames into objects and transmitting their attributes and relations instead of raw pixels.

While focusing on key content leads to compression gains, these methods' performance reduces when scene motion is fast, complex, or unpredictable
\cite{jiang2022wireless,tian2025synchronous,li2024video}. Segmentation-based schemes are particularly sensitive to such dynamics and often degrade when handling multiple moving entities. Similarly, the landmark-based SVC approach \cite{jiang2022wireless} loses subtle facial expression details, and its generalization to scenarios involving full-body motion or non-human subjects is limited. Moreover, although sending only a few keypoints is efficient, these systems must handle transmission errors in those keypoints or segmentation masks robustly, since such errors directly propagate into visual artifacts in the reconstructed video \cite{siarohin2019first,wang2018video}. These limitations highlight the need for more adaptive and resilient motion modeling in semantic compression.

The last category of research emphasizes lightweight and adaptive semantic video codecs designed for real-time operation in wireless communications. These methods aim to improve the deployability and robustness of semantic communication by making it compact, efficient, and responsive to changing channel conditions. Notable progress includes the lightweight video semantic communication (LVSC) framework \cite{li2024lvsc}, which uses a compact CNN-based optical flow estimator to extract motion vectors between frames while maintaining low complexity under varying SNR levels. Similarly, the semantic hybrid automatic repeat request (HARQ) scheme \cite{li2025semantic} addressed robustness by sending additional parity bits only when required, based on a semantic error detector that evaluates the acceptability of reconstructed frames. This selective retransmission strategy ensures high-fidelity keypoint reception while avoiding the inefficiencies of full-frame retransmission. 

Despite these advances, several limitations remain within this category. Many lightweight models, such as LVSC \cite{li2024lvsc}, are trained under specific assumptions about data and channel conditions; when real-world environments deviate from these assumptions, performance may degrade. While compact models are attractive for deployment, their reduced capacity hinders accurate modeling of high-motion or complex scenes. On the other hand, using large pre-trained generative models improves generalization but introduces substantial computational overhead, making them less practical for edge devices \cite{rombach2022high}. Additionally, mechanisms such as HARQ and feedback loops improve reliability, but often come at the cost of increased latency due to retransmissions and response delays \cite{jiang2022wireless}. 

These limits drive the development of new techniques for handling unpredictable motion and achieving robustness with efficiency. One solution is the incorporation of advanced motion prediction models from the vision perspective, such as neural motion fields (NMF) \cite{he2022nemf}. In \cite{he2022nemf}, 3D neural motion (NeMo) method was proposed that learns a global motion representation from multiple video instances of the same action. Similarly, in \cite{song2022pref}, predictability regularized neural motion field (PREF) was proposed which uses a neural field to model the motion of all points in a dynamic 3D scene. PREF imposes a regularization that penalizes chaotic, unpredictable motion, essentially encouraging the motion model to learn smooth, future-predictable trajectories. This approach was shown to handle complex scene motions, by making use of the inherent predictability in physical motion \cite{song2022pref}. Although these recent motion-aware codecs provide strong perceptual quality, they are not explicitly designed for wireless video communication and lack adaptive mechanisms for real-time model selection. Therefore in this paper, we propose PENME, which integrates prior neural motion field works (PREF \cite{song2022pref}, NeMo \cite{he2022nemf}), into wireless communication, which is developed to be a drop-in enhancement for video codecs.

PENME is designed to replace the conventional motion estimation module in video codecs with a learning-based motion field predictor that is both entropy-aware and predictability-aware. Furthermore, 
PENME models the motion of the entire scene as a continuous field using five signals, enabling it to capture both global and local transformations. Using these five signals, including motion strength, global motion consistency, peak sharpness, heterogeneity, and residual errors, PENME selects a model (CNN, optical flow, or ViT) for residual motion extraction. 
Then, on the Rx side, each frame is first reconstructed by a motion-compensated update using the transmitted residual motion. A lightweight 4-step LCM then runs only on refinement-triggered frames at reduced resolution, and the refined output is upscaled to the original size. 
Our simulation results show that by using PENME video codec, semantic-based wireless communication can achieve better compression on scenes with fast or nonlinear motion, aligning with the semantic communication goal of transmitting meaning rather than exact pixels.
\vspace{-1.5em}

\section{System Model and Problem Formulation}
\label{section:system}

In this paper, we consider a wireless video transmission framework that is designed to support Tx and Rx communication over a wireless network. Tx sends a video to Rx, in which motion information is extracted on the Tx side by PENME and encoded before transmission, in which, instead of sending full frames, Tx sends the first frame of the video and the extracted motion features between consecutive frames. The encoded motion information is then mapped onto wireless transmission packets, along with communication parameters such as bandwidth, power, and modulation scheme, ensuring adaptive resource allocation based on channel conditions, system constraints, and user quality of experience (QoE). PENME aims to minimize transmission costs by finding the most suitable model for motion extraction and also by optimizing bandwidth usage while preserving the highest possible video fidelity and semantic similarity. 


On the Rx side, the received residual motions are used to reconstruct the motion fields and regenerate the video frames. Furthermore, a diffusion-based refinement process improves residual motion for frames potentially affected by channel distortions, missing packets, or transmission errors, thus enhancing the smoothness and perceptual quality of the reconstructed video. By transmitting only residual motion data after the first frame, the system can reduce bandwidth usage while maintaining high-quality frame reconstruction.

\subsection{Overall Architecture}

The PENME framework introduces a novel approach for semantic video communication by entropy-based modality selection and predictability-aware residual motion extraction for energy-efficient and high-fidelity transmission. Instead of applying a single motion estimation method to all frames, PENME dynamically chooses between ViT, CNN, and optical flow based on the frame complexity. 
Furthermore, PENME improves motion extraction accuracy while minimizing residual data and improving compression efficiency. Additionally, PENME uses a semantics-aware policy that prioritizes residual motion updates to improve QoE and compress video efficiently in bandwidth-constrained networks.

PENME operates on consecutive frames $(F_{t-1},F_t)$ after converting them to grayscale and downscaling to $128{\times}128$ for computational efficiency. Then PENME computes five bounded signals (Eqs. \ref{eq:MS}-\ref{eq:RE}), including motion strength ($M$), global motion consistency ($C_g$), peak sharpness ($S$), heterogeneity ($H$), and residual error ($R$). These signals are fused into model scores $f_t\in\!\{\text{CNN},\text{Flow},\text{ViT}\}$ as defined in Eq. (\ref{eq:sb}), and the modality is selected using the argmax rule in Eq. (\ref{eq:modalselection}). Using the selected model, Tx estimates a robust global shift $\tilde{\Delta}_t$, and extracts a residual motion representation $R_t^{(f)}$, which, together with the modality type and minimal side information, is sent to the Rx side.

On the Rx side, \(\hat F_1\) and the side information \(\{\tilde{\Delta}_\tau,\mathcal{R}_\tau^{(f)}\}_{\tau=2}^{t}\) are available based on the video’s motion complexity and radio resources. The same modality is then applied to perform a motion-compensated update using \((\hat F_{t-1},\tilde{\Delta}_t,\mathcal{R}_t^{(f)})\) to obtain an initial estimate \(\tilde F_t\). Next, selective refinement is applied: if the residual signal \(R\) exceeds a threshold or predictability fails, a lightweight diffusion refiner (LCM-4) \(\mathcal{D}_\theta\) denoises \(\tilde F_t\); otherwise, refinement is skipped. The predictability-aware objective trains \(\mathcal{D}_\theta\) to forecast the current residual of the past frames under temporal correlation and uncertainty. Finally, the output is upscaled back to the original resolution to produce \(\hat F_t\).
\vspace{-1.2em}

\subsection{Problem Formulation in Semantic Communication}

\begin{figure*}
    \centering
    \includegraphics[width=0.937\linewidth]{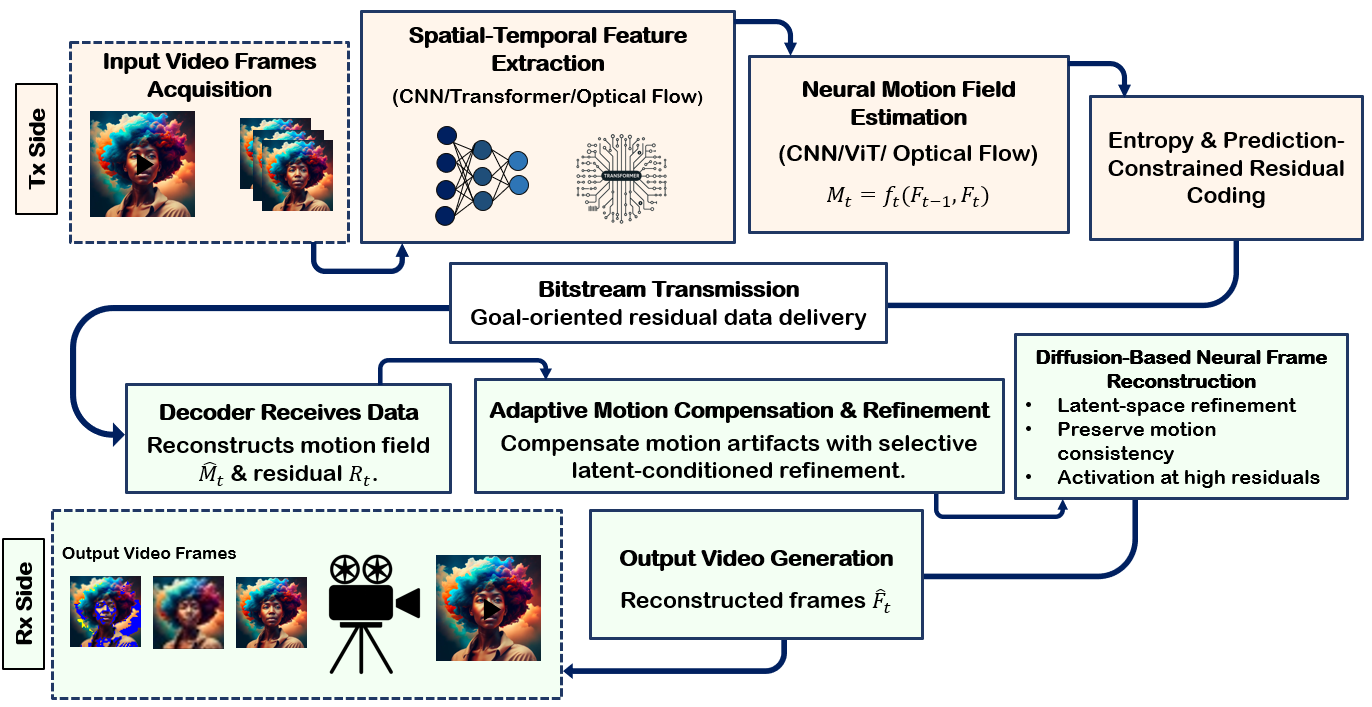}
    \caption{PENME in Video Codecs}
    \label{fig:PENME}
\end{figure*}

The optimization problem in PENME is formulated to minimize transmission costs, including the number of residual motions, bitrate, and power consumption, by ensuring accurate and fewer residual motions. In addition, PENME aims to maximize video reconstruction quality under various channel conditions, with the objective function and constraints presented in the following equations:

\begin{equation}
\Scale[0.95]{
\begin{aligned}
& \min_{M_t, R_t, B_t, P_t, Q_t^{ms}} \quad \\
& \lambda_1 C_t(M_t, R_t, B_t) + \lambda_2 P_t\quad + \lambda_3 S_t(M_t, R_t) + \lambda_4 (1 - Q_t^{ms})
\label{eqn:Fgoal}
\end{aligned}}
\end{equation}

Subject to:

\begin{equation}
\tag{\arabic{mycounter}.a}
\| M_t - \mathcal{D}_\theta(M_{t-1}, \epsilon_t) \|^2 \leq \delta,
\label{eqn:1a}
\end{equation}

\begin{equation}
\tag{\arabic{mycounter}.b}
R_t = 
\begin{cases} 
    M_t - M_{t-1}, & \text{if } \| M_t - M_{t-1} \| \geq \tau_t \text{ and } B_t > B_{\min}, \\ 
    0, & \text{otherwise}, 
\end{cases}
\label{eqn:1b}
\end{equation}

\begin{equation}
\tag{\arabic{mycounter}.c}
0 \leq B_t \leq B_{\max},
\label{eqn:1c}
\end{equation}

\begin{equation}
\tag{\arabic{mycounter}.d}
0 \leq P_t \leq P_{\max},
\label{eqn:1d}
\end{equation}

\begin{equation}
\tag{\arabic{mycounter}.e}
C_t(M_t, R_t) \leq B_t \log_2 \left( 1 + \frac{P_t h_t}{N_0} \right),
\label{eqn:1e}
\end{equation}

\begin{equation}
\tag{\arabic{mycounter}.f}
Q_t^{ms} \geq \xi_{\min},
\label{eqn:1f}
\end{equation}
where in Eq. (\ref{eqn:Fgoal}), the objective is to jointly minimize four components: the transmission rate \( C_t(M_t, R_t, B_t) \), the power consumption \( P_t \), the semantic importance cost \( S_t(M_t, R_t) \), and the multiscale structural similarity index (MS-SSIM), with respective weights \( \lambda_1, \lambda_2, \lambda_3, \lambda_4\). \( M_t \) denotes the motion field at time \( t \), and \( R_t = M_t - M_{t-1} \) is the residual motion. \( B_t \) and \( P_t \) represent the allocated RBs and the transmission power, respectively. The semantic cost term \( S_t(M_t, R_t) \) reflects adaptive motion selection based on the entropy of motion differences and quantifies the contribution of motion features and residuals to semantic or perceptual quality. \( S_t(M_t, R_t) \) balances what is transmitted with its importance for semantic understanding, particularly under bandwidth constraints. Furthermore, residuals with low semantic importance (e.g., low entropy or redundancy) are skipped, reducing unnecessary transmissions. The last component in the objective function, $Q_t^{ms}$, represents the quality of the reconstructed frames at the Rx side and directly influences the QoE. Since the objective function is defined as a minimization problem, the system uses $1-Q_t^{ms}$ to encourage improvements in reconstruction quality.

Eq. (\ref{eqn:1a}) ensures that the motion prediction error remains below the threshold \( \delta \), using a diffusion model \( \mathcal{D}_\theta \) to predict \( M_t \) from \( M_{t-1} \). This part is related to the predictability part of the PENME method as it leads to a better method for residual motion calculating. Eq. (\ref{eqn:1b}) introduces a threshold \( \tau \), which allows only significant residuals to be transmitted if enough resources \( B_t \) are available. Eqs. (\ref{eqn:1c}) and (\ref{eqn:1d}) constrain RBs and power allocation. Eq. (\ref{eqn:1e}) guarantees that transmission does not exceed Shannon channel capacity ($C_t$) based on current network conditions, where $h_t$ represents the channel gain and $N_0$ is the spectral density of the noise power. Eq. (\ref{eqn:1f}) ensures that the quality of MS-SSIM remains above a minimum threshold $\xi_{\min}$, thus directly linking optimization to the QoE of the user. This constraint guarantees perceptual quality by enforcing a lower bound on MS-SSIM. This problem formulation enables PENME to prioritize semantically important residuals under bandwidth and power constraints, improving efficiency and quality in semantic video communication. 

To solve the constrained optimization problem in Eq. (\ref{eqn:Fgoal}), we use the Lagrangian function, which transforms the constrained problem into an unconstrained one by introducing the Lagrange multipliers $\lambda$, $\mu$, and $\nu$ for the constraints. The formulation of the Lagrangian function is shown in Appendix A. Additionally, to find the optimal values of $M_t, R_t, B_t,$, $P_t$, and $Q_t^{ms}$, we apply the Karush-Kuhn-Tucker (KKT) conditions, which proofs are provided in Appendix B.

\section{PENME Residual Motion Extraction}
\label{section:performance}

The PENME approach replaces traditional motion estimation with an adaptive residual motion extraction approach. This approach aims to regenerate frames using motion vectors more accurately, reducing residual data, and improving compression efficiency. The following are the steps to design and implement PENME within a semantic video codec. Fig. \ref{fig:PENME} shows the integration of PENME which is the diagram illustrating how PENME integrates with a video transmission with the flow from video input processing to motion prediction, residual encoding, and final video reconstruction. In this scenario, the video sender is one user, and the Rx is another user.

It should be noted that PENME follows a semantic group of picture (GOP) strategy rather than frame-by-frame processing. The first frame \(F_1\) is sent as a semantic I-frame; subsequent frames are reconstructed from residual motion relative to \(F_1\). An adaptive selector (CNN/Flow/ViT) extracts residuals at the GOP level. In our 3-frame setup, \(F_3\) is reconstructed as a semantic P-frame, while \(F_2\) is skipped when its residual is below a threshold, mirroring predictive coding while reducing redundancy.

\subsection{Entropy-based Motion Feature Extraction}

Motion feature extraction has been used in semantic video transmission \cite{li2024lvsc,tong2024multimodal}, in which consecutive frames \(F_t\) and \(F_{t-1}\) are processed using a feature extractor to generate motion representations. This stage reduces redundancy and ensures that subsequent motion estimation operates on features rather than raw pixels. Previous works \cite{li2024lvsc,tong2024multimodal} relied on fixed models with manually tuned thresholds, while PENME employs a threshold-free adaptive strategy. In PENME, each frame is first converted to grayscale and down-scaled to \(128{\times}128\) to reduce redundancy and computational cost while preserving dominant motion signals. 
Then to enable model selection, five normalized motion signals are extracted to characterize motion properties. Based on these signals, the model \(f_t\in\{\text{CNN},\text{Flow},\text{ViT}\}\) is selected and its feature extractor \(f_{\vartheta}(\cdot)\) is applied to the frames.

1) \textbf{Motion Strength:}  
The overall magnitude of inter-frame changes is measured by:
\begin{equation}
    m_i = \frac{\|F_{i+1} - F_i\|_1}{\|F_{i+1}\|_1 + \|F_i\|_1 + \epsilon}, \quad 
    M = \max_i m_i,
        \label{eq:MS}
\end{equation}
where \(\|\cdot\|_1\) is the sum of the absolute pixel differences and \(\epsilon\) is a small constant added for numerical stability to avoid division by zero. 
Large values of \(M\) indicate strong motion, while small values suggest near-static scenes. Motion strength is a fundamental descriptor widely used in video activity analysis and coding \cite{ji20123d}. By construction, \(m_i\in[0,1)\) where it is zero for identical frames, and it approaches one as the differences dominate. The ratio is approximately invariant with respect to global brightness scaling. In practice, \(M\) remains close to zero for quasi-static segments and increases monotonically with motion amplitude or scene change.

2) \textbf{Global Motion Consistency:}  
Phase correlation \cite{foroosh2002extension} provides reliable estimates of translational changes between frames, as it is invariant to uniform brightness changes and robust to noise. The global motion consistency across consecutive shifts is defined as
\begin{equation}
    C_g \;=\; 1 \;-\; \frac{\sum_{i}\|\Delta_{i+1} - \Delta_i\|_2}{\sum_{i}\big(\|\Delta_i\|_2 + \|\Delta_{i+1}\|_2\big)+\epsilon},
        \label{eq:GMC}
\end{equation}
where \(C_g\!\in[0,1]\) quantifies global-motion consistency and \(\Delta_i\in\mathbb{R}^2\) is the translational shift estimated by phase correlation between frames \(F_i\) and \(F_{i+1}\). 
When the motion values $\{\Delta_i\}$ remain stable, it means that the scene is dominated by a single global transformation (such as camera panning). Conversely, variations in ${\Delta_i}$ reduce $C_g$, indicating the presence of local or irregular movements. This formulation follows the reliability measures used in video registration and global motion estimation \cite{zitova2003image,wiegand2003overview}.

3) \textbf{Peak Sharpness:}
To quantify the confidence of phase correlation for the pair $(F_i, F_{i+1})$, we compute the normalized cross-power spectrum and its correlation surface:
\begin{equation}
\Gamma_i \;=\; \frac{G_{i+1}\,\overline{G_i}}{\lvert G_{i+1}\,\overline{G_i}\rvert+\epsilon}, \qquad
r_i \;=\; \big\lvert\,\mathcal{F}^{-1}\{\Gamma_i\}\,\big\rvert,
\end{equation}
where $G_i=\mathcal{F}\{F_i\}$ and $G_{i+1}=\mathcal{F}\{F_{i+1}\}$ are 2D Fourier transforms, $\overline{G_i}$ is the complex conjugate, $\mathcal{F}^{-1}\{\cdot\}$ is the inverse Fourier transform, and $\epsilon>0$ ensures numerical stability. Then, the peak sharpness for the pair is
\begin{equation}
s_i \;=\; \frac{p_i}{p_i+\mu_i+\epsilon}\;\in(0,1),
    \label{eq:PS}
\end{equation}
where $p_i=\max r_i$ is the peak of the correlation surface and $\mu_i$ is the mean of $r_i$ over the background, computed by excluding a $(2w{+}1)\times(2w{+}1)$ neighborhood around the peak, where $w$ is the radius. 
A larger $p_i$ corresponds to a taller peak on the correlation surface, indicating stronger alignment between frames and resulting in a higher peak sharpness score $s_i$. At the sequence level, we aggregate conservatively as $S=\min_i s_i$, so that a single unreliable alignment lowers the global reliability.

4) \textbf{Heterogeneity:}
Local non-rigid dynamics are captured using block motion vectors \(\{v_j\}\) from coarse block matching \cite{barjatya2004block}. Each pair of grayscale frames is split into non-overlapping \(B\times B\) blocks; for each block, an integer pixel displacement within a \((2S{+}1)\times(2S{+}1)\) window in the next frame is chosen by minimizing the sum of absolute differences (SAD), which results in \(v_j\in\mathbb{R}^2\). The motion magnitude is \(m_j=\|v_j\|_2\), and the dispersion of \(\{m_j\}\) is summarized by a scale-normalized coefficient of variation and mapped to \([0,1)\):
\begin{equation}
\mathrm{CV}=\frac{\mathrm{std}(m_j)}{\mathrm{mean}(m_j)+\epsilon},\qquad
H=\frac{\mathrm{CV}}{1+\mathrm{CV}}\in[0,1),
    \label{eq:H}
\end{equation}
where CV is the coefficient of variation, \(\epsilon>0\) ensures numerical stability, \(B=16\) and \(S=4\), and \(\mathrm{std}(\cdot)\) and \(\mathrm{mean}(\cdot)\) are calculated over the set \(\{m_j\}\). Large \(H\) indicates heterogeneous and irregular motion (e.g., deformations or multiple moving objects), while small \(H\) reflects global uniform motion. Using CV makes \(H\) insensitive to the overall motion scale, and the squashing function \(CV/(1{+}CV)\) limits the influence of outliers.

5) \textbf{Residual Error:}
After compensating the dominant global translation, the remaining misalignment is measured by a normalized $\ell_1$ residual:
\begin{equation}
R \;=\;
\frac{\big\|\,F_{i+1}-\mathrm{warp}(F_i;\tilde{\Delta})\,\big\|_1}
{\|F_{i+1}\|_1+\|F_i\|_1+\epsilon}
\;\in[0,1),
    \label{eq:RE}
\end{equation}
where $\mathrm{warp}(F_i;\tilde{\Delta})$ applies a 2D translation to $F_i$ using bilinear interpolation with reflective padding, 
$\tilde{\Delta}=(\tilde d_x,\tilde d_y)$ is a robust global shift obtained as the componentwise median of block-matching displacements $v_j=(v_{x,j},v_{y,j})$, 
$\|\cdot\|_1$ denotes the sum of absolute pixel values.


Then, the five bounded signals are fused into three model-specific scores:
\begin{equation}
\begin{cases}
(1-M)(1-H)C_g, & \text{CNN}, \\[6pt]
M\,C_g\,S\,(1-H), & \text{Optical Flow}, \\[6pt]
M\,(1-C_g S)\,(H+R - HR), & \text{ViT}.
\end{cases}
    \label{eq:sb}
\end{equation}
where the final model is chosen by 
\begin{equation}    
f_t = \arg\max \{\text{CNN}, \text{Optical Flow},\text{ViT}\}.
    \label{eq:modalselection}
\end{equation}
It should be noted that the three candidate models are selected due to their complementary strengths:
\vspace{-0.66em}
\begin{itemize}
  \item \textbf{CNNs:} The score \((1-M)(1-H)C_g\) favors cases with low motion strength and low heterogeneity but high global consistency. In PENME, ResNet-50 is used for its efficiency and reliability; assume smooth and predictable dynamics \cite{he2016deep}. This formulation penalizes irregular or high-motion scenes, preventing CNNs from being chosen where their convolutional bias would be inadequate. The corresponding extractor $\phi_{\mathrm{CNN}}$ is calculated by the equation below:
\begin{equation}
\scalebox{0.90}{$
\begin{aligned}
\phi_{\mathrm{CNN}}(F_{t-1},F_t)
&= \mathrm{ResNet\text{-}50}\!\Big(
      \big[\,F_{t-1}\,\|\,F_t\,\|\,\big(F_t - F_{t-1}\big)\,\big]
   \Big) \\
&\in \mathbb{R}^{H' \times W' \times C},
\end{aligned}
$}
\end{equation}
  \textit{where} $[\cdot\|\cdot]$ denotes the channel concatenation, and $H',W',C$ are the spatial dimensions and channels of the ResNet feature map.
  \item \textbf{Optical flow:} The score \(M\,C_g\,S\,(1-H)\) emphasizes strong motion that is globally consistent with sharp alignment peaks, while discouraging heterogeneous fields. In PENME, Farnebäck’s method is used, which uses the strengths of dense optical flow for large, coherent displacements \cite{farneback2003two}. The corresponding extractor $\phi_{\mathrm{Flow}}$ is computed as:
\begin{equation}
\begin{aligned}
\phi_{\mathrm{Flow}}(F_{t-1},F_t)
&= \big(\mathbf{V}_t,\, \Delta_t\big), \\
\mathbf{V}_t
&= \mathrm{Farneb\ddot{a}ck}(F_{t-1},F_t) \in \mathbb{R}^{H \times W \times 2}, \\
\Delta_t
&= F_t - \mathcal{W}\!\left(F_{t-1}; \mathbf{V}_t\right),
\end{aligned}
\end{equation}
  \textit{where} $\mathbf{V}_t=(u,v)$ is the dense flow field over height $H$ and width $W$; $\mathcal{W}(\cdot;\mathbf{V}_t)$ backward-warps $F_{t-1}$ to time $t$; and $\Delta_t$ is the warp residual capturing remaining misalignment.
  \item \textbf{ViTs:} The score \(M(1-C_g S)(H+R-HR)\) is selected when long-range dependency modeling is required. PENME uses a ViViT-style encoder, where global self-attention captures non-local dependencies under complex, incoherent motion \cite{arnab2021vivit,dosovitskiy2020image}. The corresponding extractor $\phi_{\mathrm{ViT}}$ is defined as:
\begin{equation}
\begin{aligned}
\phi_{\mathrm{ViT}}(\mathcal{C}_t)
&= \mathrm{ViT\_Encoder}\!\big(\mathrm{PatchEmbed}(\mathcal{C}_t)\big) \\
&\in \mathbb{R}^{N_{\mathrm{tok}} \times d}, \\
\mathcal{C}_t
&= \{\,F_{t-L+1},\dots,F_t\,\},
\end{aligned}
\end{equation}
  \textit{where} $\mathcal{C}_t$ is a length-$L$ clip of frames; $\mathrm{PatchEmbed}(\cdot)$ partitions frames into patches and projects them to tokens; $\mathrm{ViT\_Encoder}(\cdot)$ applies spatio-temporal self-attention; $N_{\mathrm{tok}}$ is the number of tokens and $d$ the embedding dimension.
\end{itemize}
By combining bounded motion signals with model-specific strengths, PENME introduces a lightweight but powerful mechanism to extract adaptive motion characteristics. The resulting motion representations improve the accuracy of the estimate and reduce the bitrate, advancing semantic video communication for next-generation wireless systems. 

\subsection{Frame Reconstruction with Conditional Latent Refinement}

On the Rx side, PENME reconstructs each frame with the same model that was selected on the Tx side. The initial estimate is obtained by motion-compensated decoding:
\begin{equation}
\tilde F_t \;=\; \mathsf{Dec}\!\big(\hat F_{t-1}^{(\text{in})},\,\tilde{\Delta}_t,\,R_t^{(f)};\,f_t\big),
\qquad
\hat F_{t-1}^{(\text{in})} = \mathsf{gray}_{128}(\hat F_{t-1}),
\end{equation}
where $\mathsf{Dec}(\cdot)$ denotes the same–modality decoder parameterized by $f_t\!\in\!\{\text{CNN},\text{Flow},\text{ViT}\}$, $\tilde{\Delta}_t$ is the transmitted robust global shift, and $R_t^{(f)}$ is the transmitted residual representation. Then refinement is applied only to frames that are difficult to predict. A normalized residual model $R$ (Eq. \ref{eq:RE}) is mapped to its empirical cumulative distribution function (eCDF) percentile:
\begin{equation}
Rq_t \;=\; \Pr\{R \le R_t\}\in[0,1],
\end{equation}
and refinement is applied only to difficult frames:
\begin{equation}
\delta_t \;=\; \mathbf{1}\!\Big[(f_t=\text{ViT}) \;\vee\; (Rq_t \ge \tau_{\mathrm{RQ}})\Big],
\end{equation}
where $\tau_{\mathrm{RQ}}\!\in\!(0,1)$ controls selectivity (e.g., $\tau_{\mathrm{RQ}}{=}0.90$ refines only the top–residual decile). This policy aligns with the predictability-aware objective ($q(R_t\!\mid\!R_{t-1})$ and $L_{\text{PENME}}$). When the residuals are predictable, $Rq_t$ stays low and the refinement is skipped; when dynamics are uncertain, $Rq_t$ increases and selective refinement is initiated. If triggered, LCM-4 denoiser refines the estimate; otherwise the estimate is accepted:
\begin{equation}
\hat F_t^{(\text{in})} \;=\;
\begin{cases}
\mathcal{G}_\theta^{(4)}\!\big(\tilde F_t\big), & \delta_t=1,\\[2pt]
\tilde F_t, & \delta_t=0,
\end{cases}
\qquad
\hat F_t \;=\; \mathsf{up}\!\big(\hat F_t^{(\text{in})}\big),
\end{equation}
where $\mathcal{G}_\theta^{(4)}$ is a pretrained latent–consistency denoiser run for four steps, and $\mathsf{up}(\cdot)$ restores the original spatial resolution. The restored frame is mapped back to the required resolution of $428 \times 256$ pixels, and the PENME output is subsequently converted to RGB.

To avoid transmitting negligible corrections, PENME sends residuals only when their magnitude is meaningful:
\begin{equation}
T(R_t)\;=\;
\begin{cases}
\tilde R_t, & |R_t|>\epsilon,\\
0, & \text{otherwise},
\end{cases}
\end{equation}
where $\tilde R_t$ is the residual that is quantized or entropy coded and $\epsilon{>}0$ is a small threshold. Combined with the diffusion trigger $\delta_t$, this mechanism triggers refinement and transmission only when required.

It should be noted that while the mentioned considerations have been addressed, wireless data transmission is still subject to various impairments such as fading, interference, and packet loss. To maintain high similarity frame reconstruction under these conditions, PENME employs the following techniques.

\begin{itemize}
    \item \textbf{Missing Motion Compensation:} Under wireless impairments (fading, interference, packet loss), PENME uses lightweight imputation. By this consideration, if motion vectors \( M_t \) are lost due to transmission errors, the decoder estimates missing motion fields using previous motion data:
    \begin{equation}
        \Hat{M}_t = \alpha M_{t-1} + (1-\alpha) M_{t-2},
    \end{equation}
    where \( 0<\alpha<1 \) is a smoothing factor that blends past motion estimates to recover missing data. The imputed motion feeds the same update $\tilde F_t=\mathsf{Dec}(\hat F_{t-1}^{(\text{in})},\tilde{\Delta}_t,\hat M_t;f_t)$, after which the above trigger and restoration apply.

    \item \textbf{Residual Interpolation:} If the residual motion \( R_t \) is missing, the decoder estimates it using spatial interpolation:
    \begin{equation}
        R_t = \frac{1}{|\mathcal{N}|} \sum_{j \in \mathcal{N}} R_j,
    \end{equation}
    where \( \mathcal{N} \) denotes a set of neighboring motion regions and \( R_j \) are their residual values.
\end{itemize}

\subsection{Baseline: hybrid Feature Extraction}

Several works have explored hybrid feature extraction to improve video understanding. The Two-Stream convolutional networks method \cite{simonyan2014two} introduces a dual-branch CNN architecture that processes RGB frames and dense optical flow separately to capture spatial and temporal information. The optical flow guided feature method \cite{sun2018optical} improves motion representation by applying gradient operations on CNN feature maps alongside optical flow inputs. Lastly, the multimodal video transformer (MM-ViT) method \cite{chen2022mm} integrates I-frames, motion vectors, residuals, and audio using a transformer-based architecture, enabling compressed-domain semantic understanding through multi-source fusion. These approaches demonstrate the benefit of combining CNNs, optical flow, and transformers for robust video analysis.

Building on these insights, we incorporate a hybrid feature extraction baseline to assess the effectiveness of the proposed PENME method. In the hybrid method, a combination of CNN, ViT, and optical flow is considered, and for each frame, all motion residuals are extracted by all three of these methods. The extracted features are then concatenated on the Tx side, and all of that information is transmitted to the Rx side. It is evident that extracting features by three different approaches provides more detailed information about the frame and leads to more accurate frame regeneration on the Rx side by the diffusion model. In the hybrid mode, each method provides a different feature representation from the input frame \( F_t \). The unified features of different models are expressed as follows:

\[
\begin{aligned}
\mathbf{Z}_t^{\text{CNN}} &= \mathbf{W}_{\text{CNN}} \mathbf{M}_t^{\text{CNN}} \\
\mathbf{Z}_t^{\text{ViT}} &= \mathbf{W}_{\text{ViT}} \mathbf{M}_t^{\text{ViT}} \\
\mathbf{Z}_t^{\text{Flow}} &= \mathbf{W}_{\text{Flow}} \cdot \mathbf{M}_t^{\text{OpticalFlow}} 
\end{aligned}
\]

Then, we concatenate the transformed features:
\begin{equation}
\mathbf{Z}_t = \left[ \mathbf{Z}_t^{\text{CNN}}, \mathbf{Z}_t^{\text{ViT}}, \mathbf{Z}_t^{\text{Flow}} \right] \in \mathbb{R}^{d},
\end{equation}
and to capture motion dynamics over time:
\begin{equation}   
\mathbf{Z}_t^{\text{agg}} = g(\mathbf{Z}_t, \mathbf{Z}_{t-1}),
\end{equation}
where \( g(\cdot) \) is a fusion function such as concatenation. The hybrid model predicts the motion of the next frame as $\hat{\mathbf{M}}_{t+1} = f_{\text{hybrid}}(\mathbf{Z}_t^{\text{agg}})$. Then the hybrid method computes the residual motion, $\mathbf{R}_t = \mathbf{M}_t - \hat{\mathbf{M}}_t$, and this residual is encoded and transmitted.
\vspace{-0.5em}
\subsection{Baseline: Adaptive Bitrate Video Semantic Communication}

The adaptive bitrate video semantic communication (ABRVSC) method introduces a content-aware framework for efficient semantic video transmission over wireless networks \cite{gong2023adaptive}. ABRVSC uses semantic segmentation to identify regions of interest (RoIs) within each video frame. Given an input frame \( I \), a semantic segmentation network \( \mathcal{S} \) is applied to produce a semantic mask \( \mathcal{M} = \mathcal{S}(I) \), where each segmented region \( R_i \in \mathcal{M} \) corresponds to a specific semantic class. To enhance semantic understanding, a Transformer-based feature extractor is integrated to model global dependencies across the scene, enabling more accurate recognition of complex contexts. Then each region \( R_i \) is assigned an importance score based on its semantic relevance. These scores are used to determine region-specific bitrate allocations, ensuring that critical regions are encoded with higher fidelity, while background areas are more compressed by assigning lower bitrates.

The ABRVSC encoding stage includes a neural compression module that extracts a low-dimensional representation of the input frame \( I \). This representation, along with the weighted bitrate map derived from \( \mathcal{M} \), is transmitted through the wireless channel. On the Rx side, a decoder reconstructs the original frame using \( \mathcal{M} \) to guide the reconstruction process. To adapt to dynamic channel conditions, ABRVSC incorporates wireless awareness into its transmission strategy. As instantaneous SNR and total available bandwidth vary, the system dynamically adjusts bitrate allocations and transmission power to maintain efficiency under changing wireless constraints. In this paper, ABRVSC is included in the comparative study as a recent and relevant baseline that effectively integrates semantic perception with adaptive bitrate control.

\section{Results}
\label{section:results}

\begin{figure*}
    \centering
    \includegraphics[width=0.9\linewidth]{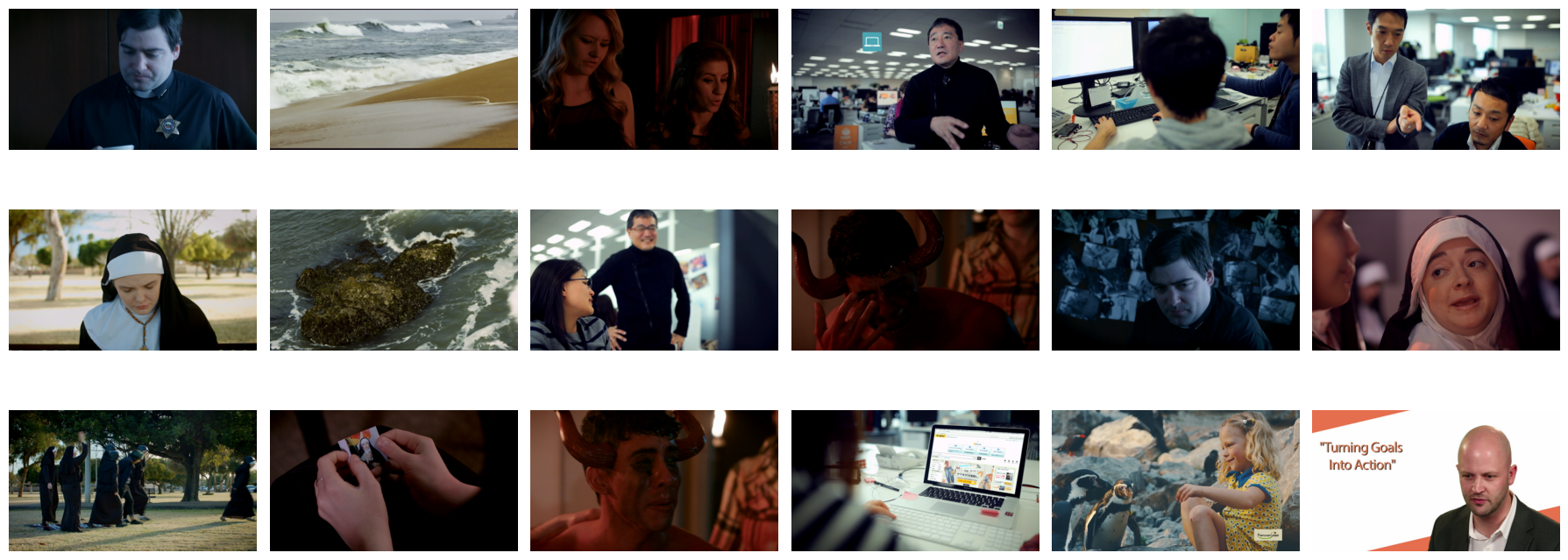}
    \caption{Samples of the Vimeo-90K dataset \cite{xue2019video}.} 
    \label{fig:DS}
\end{figure*}

\begin{figure}
    \centering
    \includegraphics[width=0.9\linewidth]{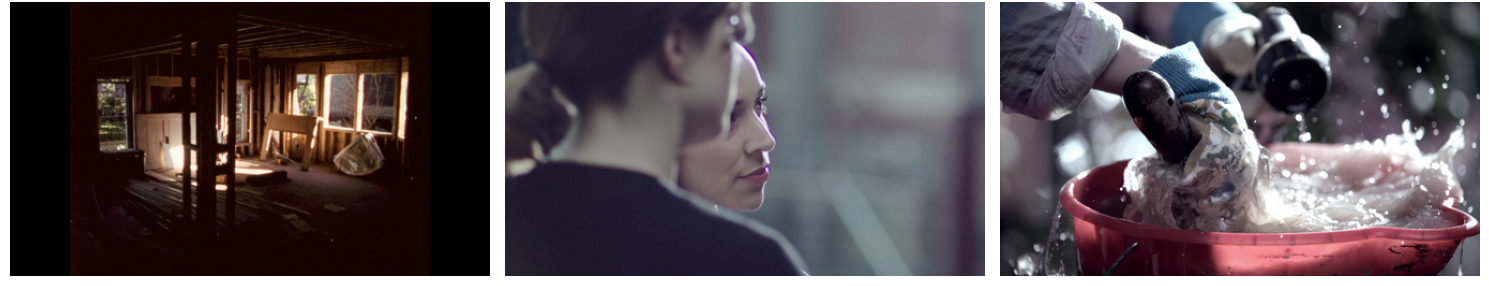}
    \caption{Samples of video frames with different entropy}
    \label{fig:sample}
\end{figure}

For evaluating PENME, we used the Vimeo-90K dataset, which provides general-purpose video sequences suitable for motion estimation. Vimeo-90K contains high-quality frame sequences with a resolution of $448 \times 256$ \cite{xue2019video} which some samples are illustrated in Fig. \ref{fig:DS}. To illustrate the variety of the dataset, we present representative frame samples from the dataset corresponding to each motion extraction technique in Fig. \ref{fig:sample}. 

Table \ref{tab:simulation_params} also summarizes the system settings used in the evaluations. All simulations and experiments in this work were implemented using Python 3.10 in the Google Colab environment. In this paper, we also assume that frames split into packets and transmitted over an orthogonal frequency division multiple access (OFDMA)-based channel, and HARQ is used for retransmissions of packets for the maximum of three times. Furthermore, additive white Gaussian noise (AWGN) and optional Rayleigh fading are also assumed. It should be noted that the modulation adaptation technique (16QAM, 64QAM, 256QAM) is used in the work to achieve better flexibility of the work against the SNR changes in the system. To assess the robustness of the system under realistic network conditions, a packet loss rate (PLR) of 2.5\% was applied in the simulation.

In this section, we present and analyze the results of PENME in comparison with ABRVSC \cite{gong2023adaptive}, traditional communication and hybrid methods in various wireless communication metrics, including energy consumption, packet drop rate (PDR), bit error rate (BER), throughput, delay, packet loss rate (PLR), and mean square error (MSE). Furthermore, we evaluate the performance of these methods using semantic communication metrics, including multiscale structural similarity (MS-SSIM), learned perceptual image patch similarity (LPIPS), peak signal-to-noise ratio (PSNR), and video quality assessment (VQA). It should be noted that the evaluation of PENME was conducted under controlled video dynamics using the Vimeo-90K dataset. This design choice allows us to isolate the algorithmic efficiency of the proposed model selection and diffusion-based refinement without the confounding factors of network-level variability.

\begin{table}[h]
    \centering
    \caption{Simulation Parameters and Hyperparameters}
    \begingroup
    \fontsize{8pt}{12pt}\selectfont
    \setlength{\tabcolsep}{6pt}
    \renewcommand{\arraystretch}{0.98}
    \begin{tabularx}{0.98\linewidth}{|l|X|}
        \hline
        \textbf{Parameter} & \textbf{Value} \\
        \hline
        \multicolumn{2}{|c|}{\textbf{General Simulation Parameters}} \\
        \hline
        Number of RBs & 10 \\
        Bandwidth per RB & 2 MHz \\
        Total Transmit Power & 1 W \\
        Noise Power Density & -174 dBm/Hz \\
        Transmission Delay & 0.1 ms \\
        Data Size per Token & 80 bits \\
        Frame Rate & 30 fps \\
        Channel Model & AWGN + Rayleigh Fading \\
        PLR & 2.5\% \\
        \hline
        \multicolumn{2}{|c|}{\textbf{PENME Methods}} \\
        \hline
        CNN & ResNet-50 \\
        ViT & Base Patch 16 \\
        optical flow & Farneback Method \\
        \(\epsilon\) & $10^{-8}$ \\
        \hline
        \multicolumn{2}{|c|}{\textbf{Diffusion Model Parameters}} \\
        \hline
        Diffusion Model Type & LCM-4 \\
        Number of Diffusion Steps & 1000 \\
        Noise Schedule & Linear \\
        Latent Representation & VQ-VAE \\
        \hline
        \multicolumn{2}{|c|}{\textbf{Hyperparameters}} \\
        \hline
        Learning Rate & 0.001 \\
        Batch Size & 32 \\
        Number of Training Epochs & 100 \\
        Loss Function & MSE \\
        Optimizer & Adam \\
        Weight Initialization & Xavier Normal \\
        \hline
    \end{tabularx}
    \endgroup
    \label{tab:simulation_params}
\end{table}

\subsection{Model Selection in PENME}

\begin{figure}
    \centering
    \includegraphics[width=0.90\linewidth]{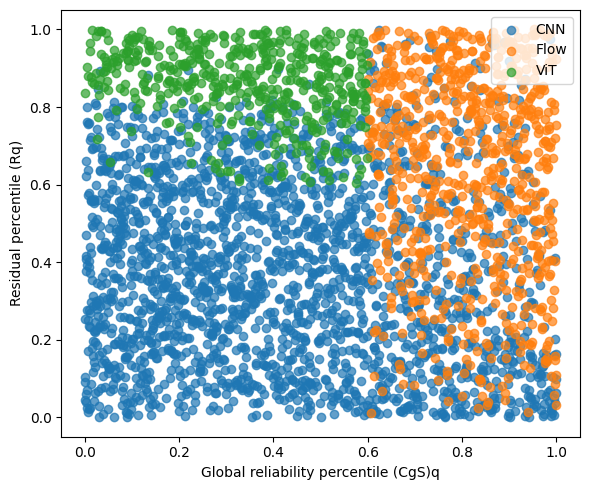}
    \caption{Entropy of the dataset video frames}
    \label{fig:Entropy}
\end{figure}

Fig. \ref{fig:Entropy} visualizes the distribution of the video frames in the global-reliability percentile \((\mathrm{CgS})_q\) versus the residual-risk percentile \(R_q\). The clusters align with PENME to select the most suitable method for residual motion extraction of the frames,
where \(R_q\) captures the expected error if no refinement is applied and \((\mathrm{CgS})_q\) summarizes the reliability of the frames. Empirically, frames with high \(R_q\) concentrate in the ViT region (upper band), those with moderate
\(R_q\) but high \((\mathrm{CgS})_q\) fall into the optical flow region (right band), and low–risk frames occupy
the CNN region (remainder), resulting in a clean decision surface.

\subsection{Energy Consumption and Total Data Size}

To compute the transmission energy required for sending data over a wireless channel, equations derived from Shannon's capacity theorem and basic energy principles are used. The channel capacity \( C \) is given by:
\begin{equation}
    C = B \cdot \log_2(1 + \text{SNR}) \cdot \zeta,
\end{equation}
where \( B \) is the channel bandwidth, \(\text{SNR}\) is on a linear scale, and $\zeta$ accounts for practical coding and modulation inefficiencies. After calculating the capacity, the transmission time \( T \) needed to send a specific amount of data is calculated as:

\begin{equation}
    T~[\text{s}] = \frac{\mathcal{D}~[\text{bits}]}{C~[\text{bits/s}]},
\end{equation}
where $\mathcal{D}$ is the number of bits to be transmitted. Then, to determine the required transmit power, the Shannon capacity equation is used to calculate the power, resulting in the expression: 

\begin{equation}
  P~[\text{W}] = \left( 2^{\frac{\text{total\_bits}~[\text{bits}]}{B~[\text{Hz}] \cdot T~[\text{s}]}} - 1 \right) \cdot \frac{N_0~[\text{W/Hz}]}{h},
\end{equation}
where \( N_0 \) is the noise power spectral density and \( h \) is the channel gain. This equation quantifies the power needed to achieve the desired data rate over a given time and bandwidth. Finally, the total transmission energy \( E \) is calculated using the relation \(E \, [\text{J}] = P \, [\text{W}] \cdot T \, [\text{s}]\). Table. \ref{tab:TDS} compares the total amount of data transmitted between three communication methods. PENME achieves a reduction in the total transmitted data, requiring approximately 110 MB, compared to 1.15 GB in traditional communication and an overwhelming 3.35 GB in hybrid method. This equals a 89.57\% reduction in data size compared to traditional communication and a 98.26\% reduction compared to the hybrid method. 

\begin{table}[ht]
\centering
\caption{Total Data Size Comparison and Improvement}
\begingroup
\fontsize{7.5pt}{10pt}\selectfont   
\setlength{\tabcolsep}{6pt}        
\renewcommand{\arraystretch}{1.05} 
\begin{tabular}{lcc}
\hline
\textbf{Method} & \textbf{Total Data Size (GB)} & \textbf{Improvement (\%)} \\
\hline
Traditional Communication & 1.15 & -- \\
PENME Method              & 0.12 & 89.57\% \\
hybrid Method  & 3.35 & -191.30\% \\
ABRVSC Method             & 0.02 & 98.26\% \\
\hline
\end{tabular}
\endgroup
\label{tab:TDS}
\end{table}

This difference is due to the design of PENME, which transmits only the residual motion information deemed semantically important, skipping redundant or low impact data. In contrast, traditional communication involves full-frame data transmission, regardless of content importance, and the hybrid method transmits the outputs of multiple models, resulting in considerable redundancy. It should be noted that the ABRVSC method exhibits the most efficient performance in terms of data volume, transmitting only around 40 MB in total, since the features of this method are extracted exclusively using transformers. Furthermore, due to the use of semantic communication for packet transmission, an additional overhead of approximately 10 bytes per packet is introduced \cite{gao2024cross, guo2024semantic}, which must be considered when optimizing system performance.

Table \ref{tab:Energy} compares the total transmitted energy between different methods, highlighting the energy efficiency of the PENME and ABRVSC methods. Traditional communication requires around $1.6\times 10^{(-9)}$ joules, whereas the hybrid semantic communication method incurs the highest energy cost at nearly $4.6\times 10^{(-9)}$ joules. The PENME method consumes approximately $1.7 \times 10^{-10}$ joules, which is significantly lower than the traditional and hybrid approaches, resulting in an 89.38\% reduction compared to traditional communication and a 96.3\% reduction compared to the hybrid method. Notably, the ABRVSC method achieves the best energy performance, consuming $1.1 \times 10^{-10}$ joules, which corresponds to a 93.13\% improvement over traditional communication. Traditional communication requires around $1.6\times 10^{(-9)}$ joules, whereas hybrid semantic communication method incurs the highest energy cost at nearly $4.6\times 10^{(-9)}$ joules.

\begin{table}[ht]
\centering
\caption{Energy consumption in the semantic and traditional communication}
\begingroup
\fontsize{8pt}{12pt}\selectfont   
\setlength{\tabcolsep}{6pt}
\renewcommand{\arraystretch}{1.05}
\begin{tabular}{lcc}
\hline
\textbf{Method} & \textbf{Total Energy (J)} & \textbf{Improvement (\%)} \\
\hline
Traditional Communication & $1.6 \times 10^{-9}$ & -- \\
PENME Method              & $1.7 \times 10^{-10}$ & 89.38\% \\
hybrid Method  & $4.6 \times 10^{-9}$  & -187.50\% \\
ABRVSC Method             & $1.1 \times 10^{-10}$ & 93.13\% \\
\hline
\end{tabular}
\endgroup
\label{tab:Energy}
\end{table}

\subsection{Packet Drop Rate and Bit Error Rate}

BER measures the fraction of bits received that are incorrect compared to the transmitted bits. In this paper, BER is calculated using the following equation.

\begin{equation}
\text{BER} = \frac{4}{\log_2 M} \cdot Q\left( \sqrt{ \frac{3 \log_2 M}{M - 1} \cdot \text{SNR}} \right),
\end{equation}

\begin{equation}
Q(x) = \frac{1}{\sqrt{2\pi}} \int_{x}^{\infty} e^{-t^2/2} \, dt = \frac{1}{2} \cdot \text{erfc}\left( \frac{x}{\sqrt{2}} \right),
\end{equation}
where $M$ is the modulation order. $Q(x)$ is the $Q$-function, representing the tail probability of the standard normal distribution, $x$ is an argument to the $Q$-function, related to the effective SNR, and $\text{erfc}(\cdot)$ is the complementary error function, defined as $\text{erfc}(x) = \frac{2}{\sqrt{\pi}} \int_x^{\infty} e^{-t^2} dt$.

PDR typically refers to the fraction of packets, frames, or data elements that are not transmitted, either intentionally (e.g., skipped due to low semantic value) or due to channel losses. PDR is calculated by the following equation:

\begin{equation}
    \text{PDR} = \frac{N_{\text{dropped}}}{N_{\text{total}}},
\end{equation}
where $N_{\text{dropped}}$ is number of skipped or lost frames or residual motions, and $N_{\text{total}}$ is the total number of data for transmission.

\begin{figure}
    \centering
    \includegraphics[width=0.91\linewidth]{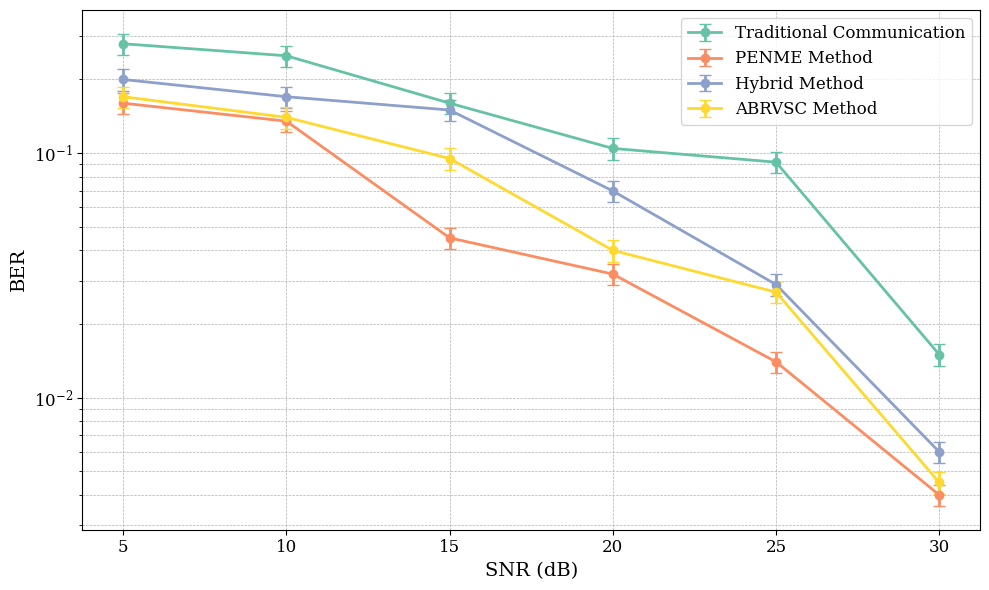}
    \caption{Bit error rate in the semantic and traditional communication}
    \label{fig:BER}
\end{figure}

Fig.~\ref{fig:BER} shows that BER decreases with SNR for all methods. PENME is consistently lowest: at 5\,dB it is around 0.20 (below hybrid/ABRVSC and far below traditional), it stays lower than hybrid and traditional at 15–20\,dB, and at 30\,dB the semantic methods drop below $10^{-2}$ while traditional remains about $1.5\times10^{-2}$. These gains arise because PENME sends a lighter and more structured residual bitstream and adapts the RBs to residual importance, which increases the energy per bit and lowers symbol error probability at the same SNR. As SNR increases, coding gains saturate and all schemes approach the channel limit, so their BERs converge.

\begin{figure}
    \centering
    \includegraphics[width=0.91\linewidth]{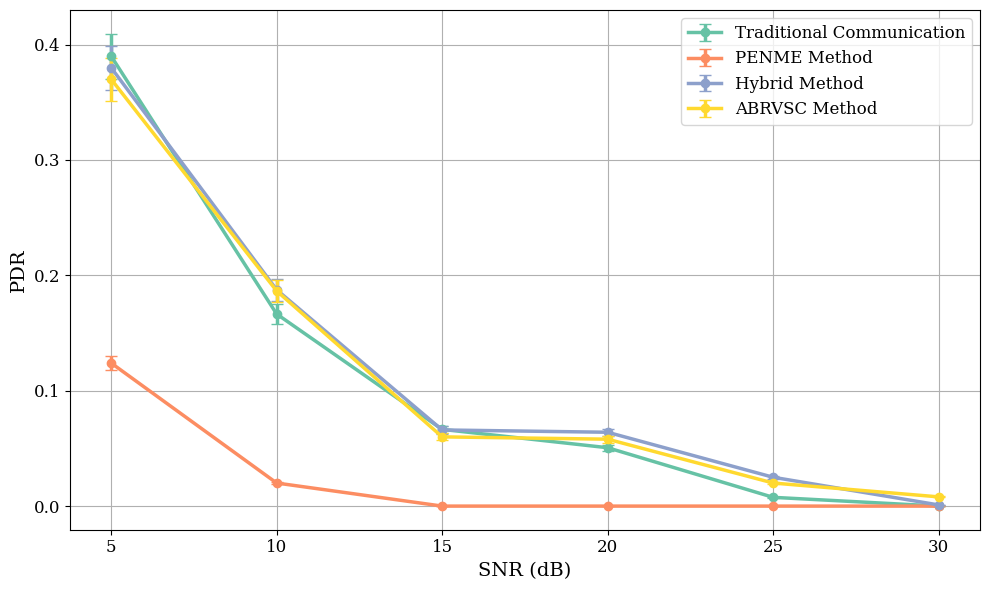}
    \caption{Packet drop rate in the semantic and traditional communication}
    \label{fig:PDR}
\end{figure}

As illustrated in Fig.~\ref{fig:PDR}, PENME has a large advantage for PDR at low SNR compared to other methods and remains lower at 10\,dB. By increasing the SNR, the PDR of the PENME decreases gradually and reaches near-zero, whereas the other methods decline more gradually. Residual-aware and channel-aware RB scheduling reduces packet count and size and lowers retransmissions. But at higher SNR, the link is reliable for all methods, so PDR differences decrease.

\subsection{Delay}

One of the most required requirements in the next generation of communication in XR services is delay, and decreasing the delay has a considerable effect on meeting the service requirements. In this paper, the delay is calculated using the following equation:

\begin{equation}
    Delay \, [\text{s}] = \frac{Total \, Time \, [\text{s}]}{Total \, Packets \,},
    \label{eq:delay}
\end{equation}

\begin{figure}
    \centering
    \includegraphics[width=0.93\linewidth]{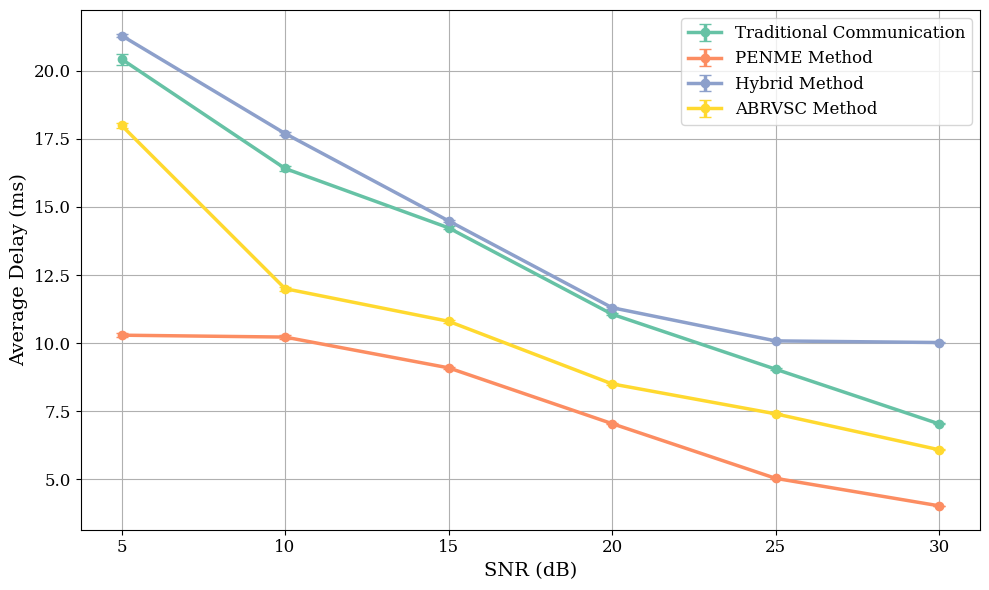}
    \caption{Delay in the semantic and traditional communication}
    \label{fig:Delay}
\end{figure}

Fig.~\ref{fig:Delay} shows the average end-to-end delay across different SNR values. The results highlight PENME's advantage in low-latency performance. At an SNR of 5\,dB, the delay for traditional and hybrid methods is approximately 20 to 21 ms, while PENME reduces this to approximately 10\,ms, achieving a 49\% improvement. As SNR increases, all methods show reduced delay due to improved channel conditions; however, PENME consistently outperforms the others. This trend continues at higher SNRs: at 20\,dB, the PENME delay is about 7\,ms, while the other two methods are similar close to 11\,ms. Finally, at 30\,dB, PENME reaches as low as 4\,ms, compared to 7\,ms and 10\,ms for traditional and hybrid methods, respectively. This substantial reduction in latency stems from PENME's efficient residual-based transmission, which avoids sending redundant frame content, and thus minimizes both packet size and transmission time. Furthermore, the ABRVSC method demonstrates competitive delay performance, closely following PENME at all SNR levels, and reaching about 6\,ms at high SNRs. By adaptively transmitting only critical semantic information, PENME significantly improves timeliness, making it applicable for real-time video applications on 5G/6G networks. Noted that the delay values reported in Fig.~\ref{fig:Delay} represent the transmission delay (Eq.~\ref{eq:delay}) and not the complete end-to-end latency. The latency also includes encoding, feature extraction, and diffusion refinement at the Rx, which incur additional computational cost depending on the hardware.

\subsection{Throughput}

\begin{figure}
    \centering
    \includegraphics[width=0.93\linewidth]{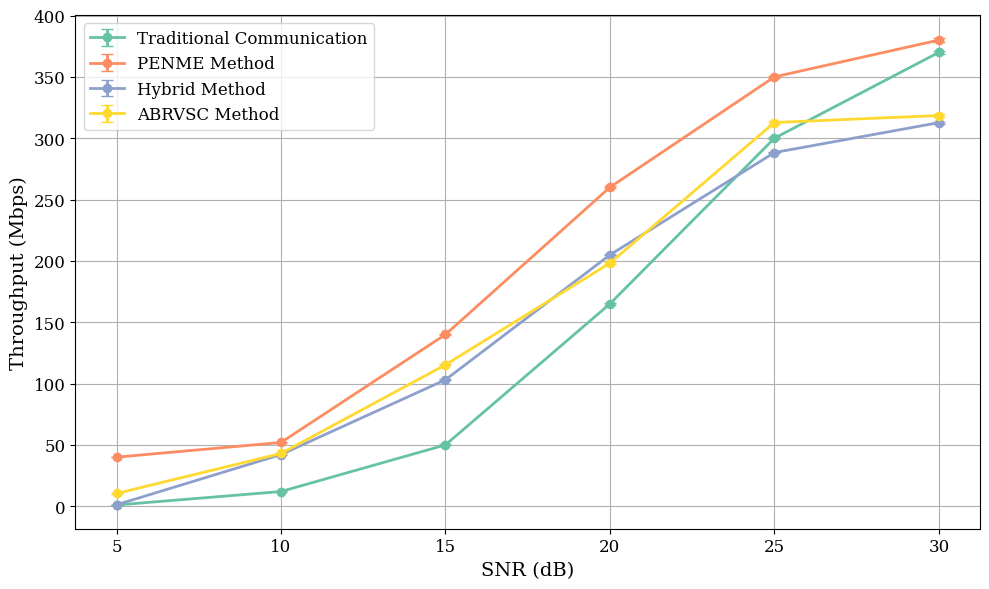}
    \caption{Throughput in the semantic and traditional communication}
    \label{fig:Throughput}
\end{figure}

Another important metric in wireless communication is throughput. Fig.~\ref{fig:Throughput} compares the throughput, which shows that PENME consistently achieves higher throughput at all SNR levels. At 5\,dB, PENME delivers approximately 42\,Mbps, outperforming ABRVSC, traditional, and hybrid methods. As SNR increases, the throughput gap widens significantly. At 15\,dB, PENME reaches 140\,Mbps, while the traditional and hybrid methods are 50\,Mbps and 105\,Mbps, respectively. By 30\,dB, PENME achieves its maximum throughput of around 385\,Mbps, compared to approximately 375\,Mbps for traditional communication, 312\,Mbps for the hybrid method, and 320\,Mbps for ABRVSC. This represents about 20\% more throughput than traditional, hybrid, and ABRVSC methods.

\subsection{Mean Square Error (MSE)}

All the metrics discussed so far are network-level indicators, demonstrating that PENME outperforms existing methods in those aspects. However, PENME is designed as a semantic communication method and should also provide better performance in semantic-specific metrics. MSE measures the pixel-wise difference between two frames and shows distortion in frame reconstruction. In this paper, MSE is calculated using the following equation: 

\begin{equation}
\mathrm{MSE}(F_t,\hat{F}_t)
= \frac{1}{|\mathcal{P}|}\sum_{p\in\mathcal{P}} \big(F_t(p)-\hat{F}_t(p)\big)^2,
\end{equation}
\noindent
where $F_t$ is the ground-truth frame at time $t$, $\hat{F}_t$ is the reconstructed frame,
$\mathcal{P}$ is the set of pixel locations, and $|\mathcal{P}|=H\times W$.

\begin{figure}
    \centering
    \includegraphics[width=0.93\linewidth]{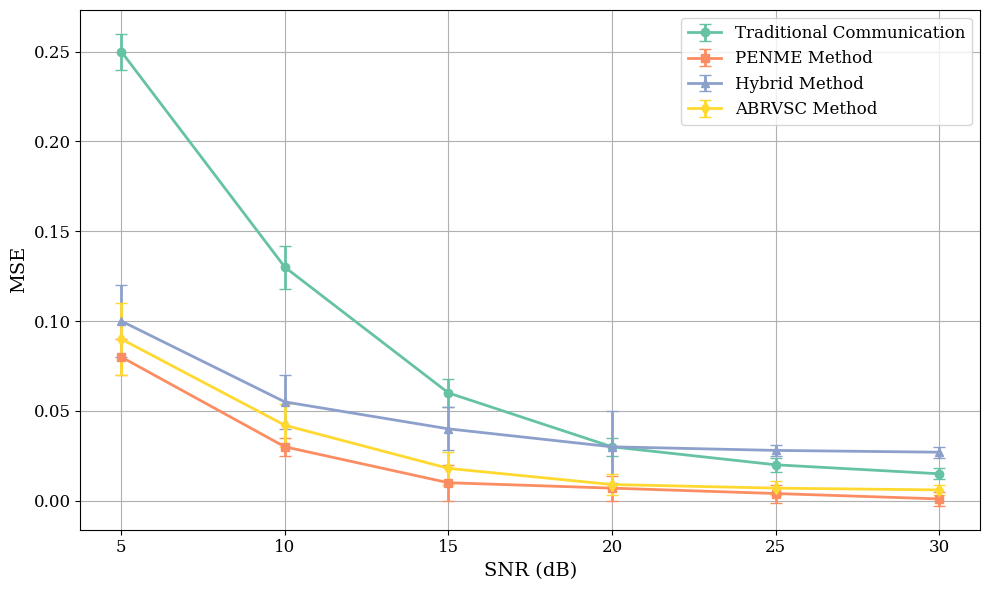}
    \caption{MSE in the semantic and traditional communication}
    \label{fig:MSE}
\end{figure}

Fig.~\ref{fig:MSE} presents the performance of MSE in which the PENME method consistently achieves the lowest MSE in all SNR values, indicating its superior reconstruction quality. At a low SNR of 5~dB, traditional communication has the highest error at around 0.252, while ABRVSC, hybrid, and PENME methods achieve much lower errors of approximately 0.095, 0.103, and 0.080, respectively. This early advantage of semantic communication becomes more pronounced at higher SNRs. As the SNR increases to 30~dB, PENME reaches an MSE as low as 0.0015, outperforming both hybrid (0.025) and traditional (0.012) methods. The performance gap between PENME and the other methods demonstrates its capability to minimize pixel-wise reconstruction errors, especially in noisy or low-resource scenarios.

The results show that adaptive method selection is an effective solution, as PENME achieves the lowest MSE across all SNR levels due to its entropy- and predictability-aware residual motion estimation, which prioritizes encoding only the most critical semantic changes. This targeted compression strategy results in a lower compression cost while preserving essential motion features. In contrast, the hybrid method, which transmits features from multiple models, introduces fusion noise and lacks effective redundancy control, making it more susceptible to errors, especially under low SNR conditions. Traditional communication, which transmits full video frames without semantic awareness or compression, suffers from the highest MSE due to its inability to filter irrelevant information or adapt to channel variability.

\subsection{Peak Signal-to-Noise Ratio (PSNR)}

PSNR is also a widely used metric to assess image reconstruction quality, especially after transmission or compression. PSNR measures the difference between the original and reconstructed frames, and a higher PSNR means better fidelity. PSNR evaluates the visual quality of the reconstructed video on the Rx side. In this paper, PSNR is calculated using the following equation:
\begin{equation}
    PSNR = 10\cdot \log \frac{(I_{max}^2)}{MSE},
\end{equation}
where $I_{max}$ is the maximum possible pixel value, which is equal to one for normalized images.

\begin{figure}
    \centering
    \includegraphics[width=0.93\linewidth]{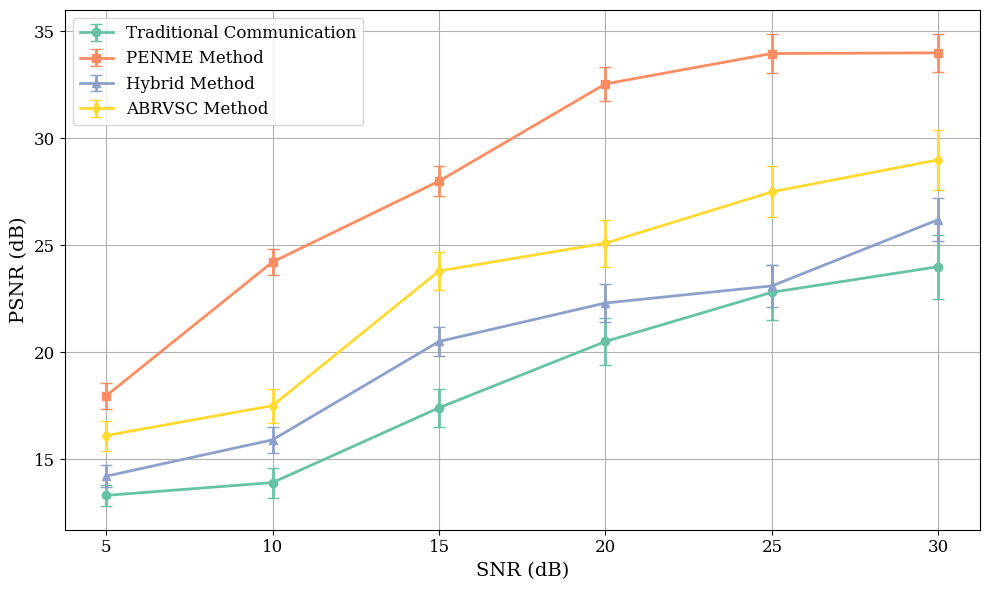}
    \caption{PSNR in the semantic and traditional communication}
    \label{fig:PSNR}
\end{figure}

Fig.~\ref{fig:PSNR} illustrates the reconstruction quality of the received video frames at different SNR levels. As expected, all methods improve as the SNR increases; however, PENME consistently outperforms the baselines. At a low SNR of 5\,dB, traditional communication is approximately 13.3\,dB, while the hybrid and ABRVSC methods reach roughly 14.5\,dB and 16.0\,dB, and PENME attains about 18.2\,dB. As SNR increases to 15\,dB, PENME increases to about 28.3\,dB, exceeding ABRVSC, the hybrid method and traditional communication. This trend continues at higher SNRs; at 30\,dB, PENME reaches 34 dB, exceeding traditional communication by 10 dB and remaining above ABRVSC 29.0 dB and the hybrid method 26.3 dB.

PENME’s PSNR performance is a result of its adaptive residual motion estimation, which emphasizes choosing semantically relevant regions while filtering out noise and redundancy. In contrast, traditional communication, lacking any semantic filtering or compression, transmits raw pixel data that degrades under noisy conditions. The hybrid method also performs well, but the fusion of heterogeneous features can have some detrimental effects, and there may be conflicts between modalities during the extraction of features from frames. Overall, these results indicate that while PENME is slightly less energy efficient than ABRVSC, it offers substantial energy savings and benefits from better overall performance in other metrics such as delay, MSE, and PSNR.

\subsection{MS-SSIM}

MS-SSIM is an extension of the standard SSIM metric designed to evaluate perceptual similarity between $F_t$ and its reconstruction $\hat{F}_t$ at multiple spatial resolutions. Unlike pixel-wise error metrics, such as MSE, MS-SSIM models the human visual system by comparing luminance, contrast, and structural information across progressively down-sampled versions of the images. The MS-SSIM value is obtained by combining these measurements at different scales, making it more robust to distortions and better aligned with perceived quality. Formally, the metric is expressed as:
\begin{equation}
\begin{aligned}
\mathrm{MS\text{-}SSIM}(F_t,\hat{F}_t)
&= \left[l_M(F_t,\hat{F}_t)\right]^{\alpha_M} 
   \cdot \prod_{j=1}^{M} \left[c_j(F_t,\hat{F}_t)\right]^{\beta_j} \\
&\quad \cdot \left[s_j(F_t,\hat{F}_t)\right]^{\gamma_j},
\end{aligned}
\end{equation}
where $l_M$, $c_j$, and $s_j$ denote the luminance, contrast, and structure comparisons at scale $j$, with exponents $\alpha_M$, $\beta_j$, and $\gamma_j$ controlling their relative importance. The resulting score ranges from zero to one, with higher values indicating stronger structural similarity and thus better perceptual quality.

\begin{figure}
    \centering
    \includegraphics[width=0.93\linewidth]{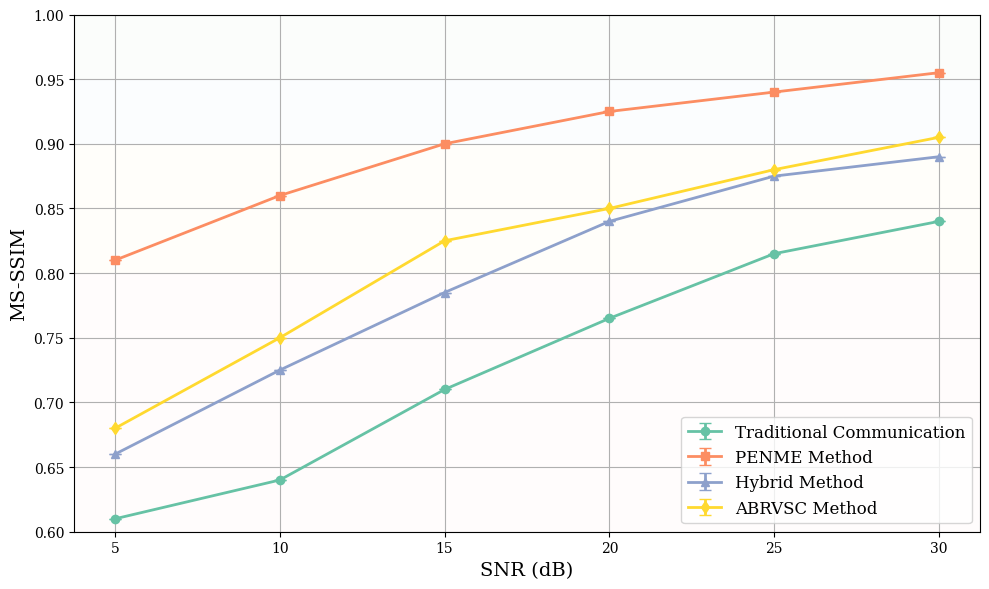}
    \caption{MS-SSIM in the semantic and traditional communication}
    \label{fig:MSSSIM}
\end{figure}

Fig.~\ref{fig:MSSSIM} illustrates the performance of MS-SSIM as a function of SNR. It is evident that the PENME method consistently provides the highest perceptual quality across all SNR values, achieving an MS-SSIM of approximately 0.81 at 5~dB and rising steadily to about 0.96 at 30~dB. In comparison, traditional communication starts much lower at around 0.61 and only reaches 0.84 at high SNR. Both ABRVSC and hybrid methods perform in between, with ABRVSC slightly exceeding hybrid, especially at mid-to-high SNR regions. These results confirm that semantic-oriented methods, particularly PENME, can maintain structural similarity much more effectively under noisy conditions, while traditional approaches struggle to capture perceptual consistency.

\subsection{LPIPS}

LPIPS is based on deep feature extraction, where both  $F_t$ and its reconstruction $\hat{F}_t$ are passed through a pretrained CNN $\phi(\cdot)$. The distance is then calculated as
\begin{equation}
\begin{aligned}
\mathrm{LPIPS}(F_t,\hat{F}_t)
&= \sum_{l} \frac{1}{H_l W_l} \sum_{h=1}^{H_l} \sum_{w=1}^{W_l} \\
&\quad \left\|\, w_l \odot \big(\phi_l(F_t)_{h,w} - \phi_l(\hat{F}_t)_{h,w}\big) \right\|_2^2 ,
\end{aligned}
\end{equation}
where $l$ indexes the feature layers, $H_l \times W_l$ are the spatial dimensions, and $w_l$ are learned weights that calibrate perceptual relevance. Lower LPIPS values indicate stronger perceptual similarity.

\begin{figure}
    \centering
    \includegraphics[width=0.93\linewidth]{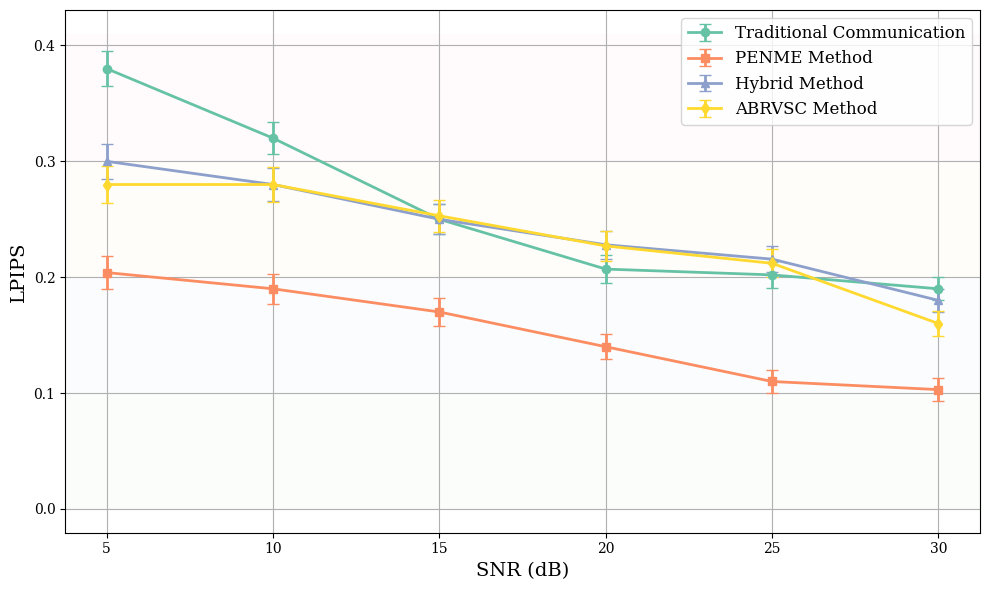}
    \caption{LPIPS in the semantic and traditional communication}
    \label{fig:LPIPS}
\end{figure}

Fig.~\ref{fig:LPIPS} shows the LPIPS performance in different SNR values. As expected, lower values of LPIPS indicate higher perceptual similarity, and the PENME method consistently achieves the lowest scores across the entire SNR range. At 5~dB, PENME achieves an LPIPS of approximately 0.20, while traditional communication, hybrid, and ABRVSC methods remain higher at approximately 0.38, 0.30, and 0.28, respectively. As SNR increases, all methods demonstrate improvement, but PENME maintains a clear margin, reaching as low as 0.10 at 30~dB. In contrast, other semantic-based methods (hybrid and ABRVSC) converge around 0.18 and 0.16, while traditional communication still lags slightly behind at 0.19. These results emphasize that PENME is more effective in preserving perceptual similarity under both noisy and high-quality conditions, showing the strongest alignment with human visual perception.

\begin{figure}
    \centering
    \includegraphics[width=0.90\linewidth]{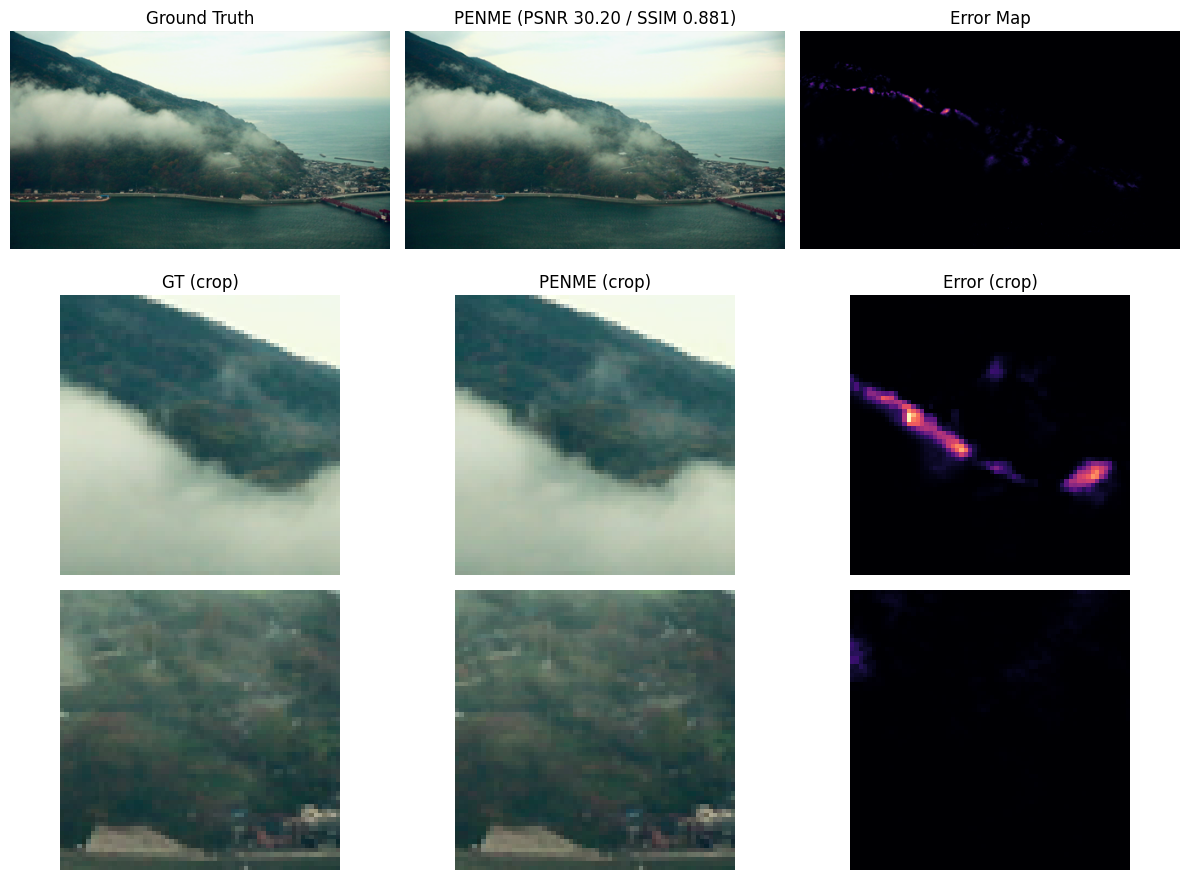}
    \caption{PENME performance on a dataset sample}
    \label{fig:res}
\end{figure}

Fig. \ref{fig:res} presents the reconstructed video frame generated by PENME against the ground-truth frame. The error map highlights localized distortions, primarily along structural edges. The cropped patches show that PENME preserves fine textures and structural consistency, achieving high perceptual quality with $PSNR = 30.20$ dB and $SSIM = 0.881$. The error visualization confirms that residual errors are minor and concentrated in high-frequency areas, demonstrating the effectiveness of PENME in maintaining semantic and visual fidelity.

\section{Discussion}
\label{discussion}

The proposed PENME framework introduces a novel approach to semantic video communication by integrating predictability- and entropy-aware residual motion extraction on the Tx side and latent-space refinement at the Rx, which then guides frame reconstruction. The refinement stage uses LCM-4, a lightweight latent consistency model that performs refinement in only four denoising steps, significantly reducing the number of iterations compared to conventional diffusion methods. Experimental results show that PENME significantly reduces bitrate and energy consumption while maintaining high semantic similarity and reconstruction quality. Another key aspect of PENME is that ResNet-50, ViViT, and optical flow are used solely for feature extraction and residual motion estimation, without requiring pretraining or fine-tuning. This design makes the framework model-agnostic, since these components only provide motion-related features rather than being used as full prediction models. Additionally, all video frames are resized to a resolution of $128\times128$ and converted to grayscale before processing, which reduces computational load and memory usage while preserving motion-related information.

While this design ensures flexibility and efficiency, it still assumes that residual motion can be accurately predicted and transmitted sparsely. This assumption may not hold in videos with highly irregular motion patterns or unpredictable frame dynamics, where prediction errors could accumulate and degrade the reconstruction quality. The overall computational cost of PENME can be divided into three main components:

\begin{itemize}
    \item \textbf{Entropy and Predictability Estimation:} For each frame, entropy is calculated, and the prediction quality is estimated using the selected model (CNN, ViT, or optical flow). This step has complexity $O(n)$, where $n$ is the number of pixels after resizing.
    \item \textbf{Feature Extraction and Residual Computation:} Based on the selected model, optical flow has complexity $O(n\log n)$, CNN-based extraction has complexity $O(nd)$, where $d$ is the number of network layers (49 in ResNet-50, as the fully connected layer is not used), and ViT-based extraction can reach $O(n^2)$ due to self-attention.
    \item \textbf{Diffusion-Based Frame Reconstruction:} At the decoder, the refinement process using LCM-4 requires only $T=4$ iterations, each with complexity depending on model depth $d$ and image resolution, typically $O(nd)$ per step. This reduces the total cost to $O(4nd)$ compared to $O(Tnd)$ with large $T$.
\end{itemize}

Although these components remain manageable on GPUs, they may pose challenges in constrained environments. To address these challenges, future work could explore the selection of edge-aware and context-driven models to further reduce runtime while preserving perceptual quality. Furthermore, we acknowledge that real-world wireless environments introduce additional challenges, such as bandwidth fluctuations, mobility, and XR-specific delay constraints. Addressing this dimension is an important next step. In future work, we plan to couple PENME with established 5G/6G link-level simulators or live-trace repositories, enabling full end-to-end evaluation under realistic channel dynamics.

\section{Conclusion}
\label{section:conclusion}
In this paper, we propose PENME, a semantic communication framework for video transmission in wireless networks. PENME introduces an adaptive residual motion extraction approach that uses entropy-based modality selection combined with predictability-aware correction, dynamically selecting between optical flow, CNN, and ViT to better match frame complexity. PENME decodes with the same model branch as Tx in a compact working domain, selectively refines only hard frames in latent space, and restores frames to their original resolution, achieving high fidelity while controlling bitrate and computation. The proposed PENME system formulates an optimization problem that balances bitrate, power, and semantic importance under wireless resource constraints. To address the computational complexity of exact optimization, PENME employs a practical adaptive filtering mechanism based on entropy thresholds and predictability scores. Extensive experiments demonstrate that PENME provides consistent improvements in semantic video reconstruction, compression efficiency, and transmission reliability under varying network conditions.

\section*{Acknowledgment}

This work has been supported by MITACS, Ericsson Canada, and partially by NSERC Collaborative Research and Training Experience Program (CREATE) TRAVERSAL under Grant 497981 and Canada Research Chairs program. 

\bibliographystyle{IEEEtran}
\bibliography{references}

\appendix
\renewcommand{\thesection}{\Alph{section}.\arabic{section}}
\setcounter{section}{0}

\begin{appendices}

\section*{A. Constructing the Lagrangian Function}

To solve the constrained optimization problem in Eq. \ref{eqn:Fgoal}, we use the Lagrangian function, which transforms the constrained problem into an unconstrained one by introducing Lagrange multipliers $\lambda$, $\mu$, and $\nu$ for the constraints. The Lagrangian function is formulated as:
\begin{multline}
\mathcal{L} = \lambda_1 C_t(M_t, R_t, B_t) + \lambda_2 P_t 
+ \lambda_3 S_t(M_t, R_t) + \lambda_4 (1 - Q_t^{ms}) \\
+ \mu_1 (|| M_t - \mathcal{D}_\theta(M_{t-1}, \epsilon_t) ||^2 - \delta) \\
+ \mu_2 (C_t(M_t, R_t) - B_t \log_2 (1 + \tfrac{P_t h_t}{N_0})) 
+ \mu_3 (\xi_{\min} - Q_t^{ms}).
\label{eqn:Lagrangian}
\end{multline}
which covers the objective function, ensures the motion estimation accuracy constraint, enforces the bandwidth constraint using the Shannon capacity formula, and guarantees perceptual quality by enforcing the MS-SSIM threshold.
By solving $\frac{\partial \mathcal{L}}{\partial M_t} = 0$, $\frac{\partial \mathcal{L}}{\partial R_t} = 0$, $\frac{\partial \mathcal{L}}{\partial B_t} = 0$, $\frac{\partial \mathcal{L}}{\partial P_t} = 0$, and $\frac{\partial \mathcal{L}}{\partial Q_t^{ms}} = 0$, we obtain the necessary conditions for an optimal solution.

\section*{B. Solving the Optimization Problem Using KKT Conditions}

To find the optimal values of $M_t, R_t, B_t,$, $P_t$, and $Q_t^{ms}$ we apply the Karush-Kuhn-Tucker (KKT) conditions, which provide necessary conditions for an optimal solution under constraints. For meeting the requirements of KKT, the following conditions should be met:
\begin{itemize}
    \item Stationarity: The gradient of the Lagrangian function should be zero:
    \[
   \frac{\partial \mathcal{L}}{\partial M_t} = 0, \quad
   \frac{\partial \mathcal{L}}{\partial R_t} = 0, \quad
   \frac{\partial \mathcal{L}}{\partial B_t} = 0, \quad
   \label{eqn:KKT1}
   \]

    \[
   \frac{\partial \mathcal{L}}{\partial P_t} = 0, \quad
    \frac{\partial \mathcal{L}}{\partial Q_t^{ms}} = 0.
   \label{eqn:KKT2}
   \]
   
   \item     Primal Feasibility: The constraints should be satisfied:
   \[
   || M_t - \mathcal{D}_\theta(M_{t-1}, \epsilon_t) ||^2 \leq \delta, \quad \]
   \[
   C_t(M_t, R_t) \leq B_t \log_2 \left( 1 + \frac{P_t h_t}{N_0} \right), \quad\
   \]
    \[
    Q_t^{ms} \geq \xi_{\min}.
    \]
   \item Dual Feasibility: The Lagrange multipliers must be non-negative:
   \[
   \mu_1 \geq 0, \quad \mu_2 \geq 0, \quad \mu_3 \geq 0.
   \]

   \item Complementary Slackness: The product of each constraint and its corresponding Lagrange multiplier should be zero:
   \begin{multline}
   \mu_1 (|| M_t - \mathcal{D}_\theta(M_{t-1}, \epsilon_t) ||^2 - \delta) = 0, \quad \\
   \mu_2 (C_t(M_t, R_t) - B_t \log_2 (1 + \frac{P_t h_t}{N_0} )) = 0,  \\  
   \mu_3 (\xi_{\min} - Q_t^{ms}) = 0.
   \label{eqn:KKT4}
   \end{multline}
\end{itemize}

By solving the system of equations given by KKT conditions, we derive the optimal allocation of motion fields, residuals, bandwidth, power, and perceptual quality (MS-SSIM).

\section*{C. optical flow Estimation}

Expand a function as an infinite sum of its derivatives:

\[
f(x + \delta x) = f(x) + \frac{\partial f}{\partial x} \delta x + \frac{\partial^2 f}{\partial x^2} \frac{\delta x^2}{2!} + \dots + \frac{\partial^n f}{\partial x^n} \frac{\delta x^n}{n!},
\]
in which if \( \delta x \) is small:

\[
f(x + \delta x) = f(x) + \frac{\partial f}{\partial x} \delta x + O(\delta x^2),
\]
where \( O(\delta x^2) \) is almost zero.

For a function of three variables with small \( \delta x, \delta y, \delta t \):

\[
f(x + \delta x, y + \delta y, t + \delta t) \approx f(x, y, t) + \frac{\partial f}{\partial x} \delta x + \frac{\partial f}{\partial y} \delta y + \frac{\partial f}{\partial t} \delta t
\]
where, \(k\) and \(l\) denote discrete spatial indices for pixel location along the horizontal (column) and vertical (row) directions, respectively, while \(t\) represents the temporal frame index.

Using central differences to approximate image gradients, the discrete derivative along the $x$-direction is given as:

\begin{multline}
I_x(k,l,t) =\\
\frac{1}{4} \big[ I(k+1,l,t) + I(k+1,l,t+1) +\\ I(k+1,l+1,t) + I(k+1,l+1,t+1) \big] - \\
\frac{1}{4} \big[I(k-1,l,t) + I(k-1,l,t+1) + \\ I(k-1,l+1,t) + I(k-1,l+1,t+1) \big]
\end{multline}

Similarly, $I_y(k,l,t)$ and $I_t(k,l,t)$ can be obtained using central differences in the $y$- and $t$-directions, respectively.

\end{appendices}

\end{document}